\documentclass[10pt, onecolumn]{IEEEtran}
\usepackage{cite}
\usepackage{hyperref}
\usepackage{amsthm}
\usepackage{amsmath}
\usepackage{amsfonts}
\usepackage{graphicx}
\usepackage{latexsym}
\usepackage{amssymb}
\usepackage{stmaryrd}
\usepackage{comment}
\usepackage{amscd}
\usepackage[small]{caption}
\usepackage{algorithmicx}
\usepackage{algorithm}
\algrenewcommand\alglinenumber[1]{\scriptsize #1:}
\algrenewcommand\algorithmicindent{1em}%
\allowdisplaybreaks
\usepackage{graphicx}
\usepackage{subfigure}
\usepackage{wrapfig}
\usepackage{float}
\usepackage{xcolor}
\newcommand{\sgn}{\text{sgn}}

\usepackage{mathtools}
\newcommand{\bea}{\begin{eqnarray}}
\newcommand{\eea}{\end{eqnarray}}
\newcommand{\ceil}[1]{\left\lceil #1 \right\rceil}
\newcommand{\floor}[1]{\left\lfloor #1 \right\rfloor}

\newcommand{\sbinom}[2]{\left( \begin{array}{c} #1 \\ #2 \end{array} \right) }

\newcommand{\cA}{{\cal A}}

\newcommand{\cF}{{\cal F}}

\newcommand{\cL}{{\cal L}}

\newcommand{\cR}{{\cal R}}
\newcommand{\cS}{{\cal S}}

\newcommand{\cW}{{\cal W}}
\newcommand{\cX}{{\cal X}}
\newcommand{\cY}{{\cal Y}}

\newcommand{\sG}{\script{G}}

\newcommand{\sP}{\script{P}}

\newcommand{\bfg}{{\boldsymbol g}}

\newcommand{\bfx}{{\boldsymbol x}}
\newcommand{\bfy}{{\boldsymbol y}}
\newcommand{\bfz}{{\boldsymbol z}}

\DeclareMathOperator*{\argmax}{arg\,max}
\DeclareMathOperator*{\argmin}{arg\,min}

\newcommand{\kDel}{k\textrm{-}\mathsf{Del}}
\newcommand{\kIns}{k\textrm{-}\mathsf{Ins}}

\DeclareMathAlphabet{\mathbfsl}{OT1}{cmr}{bx}{it}
\newcommand{\uuu}{\kern-1pt\mathbfsl{u}\kern-0.5pt}
\newcommand{\vvv}{\kern-1pt\mathbfsl{v}\kern-0.5pt}

\newcommand{\myboxplus}{\kern1pt\mbox{\small$\boxplus$}}

\makeatletter \DeclareRobustCommand{\sbinom}{\genfrac[]\z@{}}
\makeatother
\newcommand{\G}[2]{\sbinom{{#1}\kern-1pt}{{#2}\kern-1pt}}
\newcommand{\Gq}[2]{\sbinom{{#1}\kern-0.25pt}{{#2}\kern-0.25pt}}

\newcommand{\Ps}{\smash{{\sP\kern-2.0pt}_q\kern-0.5pt(n)}}
\newcommand{\sPs}{\smash{{\sP\kern-1.5pt}_q(n)}}
\newcommand{\Ptwo}{\smash{{\sP\kern-2.0pt}_2\kern-0.5pt(n)}}
\newcommand{\Ptwom}{\smash{{\sP\kern-2.0pt}_2\kern-0.5pt(m)}}
\newcommand{\Ptwonm}{\smash{{\sP\kern-2.0pt}_2\kern-0.5pt(n+m)}}
\newcommand{\Ptwoa}{\smash{{\sP\kern-2.0pt}_2\kern-0.5pt(1)}}
\newcommand{\Ptwob}{\smash{{\sP\kern-2.0pt}_2\kern-0.5pt(2)}}
\newcommand{\Ptwoc}{\smash{{\sP\kern-2.0pt}_2\kern-0.5pt(3)}}
\newcommand{\Ptwod}{\smash{{\sP\kern-2.0pt}_2\kern-0.5pt(4)}}
\newcommand{\Ptwoe}{\smash{{\sP\kern-2.0pt}_2\kern-0.5pt(5)}}
\newcommand{\Ptwof}{\smash{{\sP\kern-2.0pt}_2\kern-0.5pt(6)}}
\newcommand{\Ptwokm}{\smash{{\sP\kern-2.0pt}_2\kern-0.5pt(2k-1)}}
\newcommand{\Pone}{\smash{{\sP\kern-2.5pt}_2\kern-0.5pt(n{-}1)}}

\newcommand{\Gr}{\smash{{\sG\kern-1.5pt}_q\kern-0.5pt(n,k)}}
\newcommand{\Gi}{\smash{{\sG\kern-1.5pt}_q\kern-0.5pt(n,i)}}
\newcommand{\Gj}{\smash{{\sG\kern-1.5pt}_q\kern-0.5pt(n,j)}}
\newcommand{\Grmk}{\smash{{\sG\kern-1.5pt}_q\kern-0.5pt(n,n-k)}}
\newcommand{\Grdk}{\smash{{\sG\kern-1.5pt}_q\kern-0.5pt(2k,k)}}
\newcommand{\Grekappa}{\smash{{\sG\kern-1.5pt}_q\kern-0.5pt(n,e+1-\kappa)}}
\newcommand{\Grtwoekappa}{\smash{{\sG\kern-1.5pt}_q\kern-0.5pt(n,2e+1-\kappa)}}
\newcommand{\Gremkappa}{\smash{{\sG\kern-1.5pt}_q\kern-0.5pt(n,e-\kappa)}}
\newcommand{\Gn}{\smash{{\sG\kern-1.5pt}_2\kern-0.5pt(n,n{-}1)}}
\newcommand{\Gnq}{\smash{{\sG\kern-1.5pt}_q\kern-0.5pt(n,n{-}1)}}
\newcommand{\Gone}{\smash{{\sG\kern-1.5pt}_2\kern-0.5pt(n,1)}}
\newcommand{\Gqone}{\smash{{\sG\kern-1.5pt}_q\kern-0.5pt(n,1)}}
\newcommand{\GTwo}{\smash{{\sG\kern-1.5pt}_2\kern-0.5pt(n,k)}}
\newcommand{\GTwonk}[2]{{\smash{{\sG\kern-1.5pt}_2\kern-0.5pt({#1},{#2})}}}
\newcommand{\Gnk}{\smash{{\sG\kern-1.5pt}_2\kern-0.5pt(n,n{-}k)}}
\newcommand{\Greone}{\smash{{\sG\kern-1.5pt}_q\kern-0.5pt(n,e{+}1)}}
\newcommand{\Gretwo}{\smash{{\sG\kern-1.5pt}_q\kern-0.5pt(n,e{+}2)}}

\newcommand{\be}[1]{\begin{equation}\label{#1}}
\newcommand{\ee}{\end{equation}}

\newcommand{\Cref}[1]{Co\-rol\-la\-ry\,\ref{#1}}

\newtheorem{theorem}{Theorem}
\newtheorem{lemma}{Lemma}
\newtheorem{remark}{Remark}
\newtheorem{corollary}{Corollary}
\newtheorem{claim}{Claim}

\newtheorem{problem}{Problem}

\begin{document}

 \author{%
   \IEEEauthorblockN{{Shubhransh~Singhvi}\IEEEauthorrefmark{1}, {Omer~Sabary}\IEEEauthorrefmark{2}, 
                     {Daniella~Bar-Lev}\IEEEauthorrefmark{2}
                     and {Eitan~Yaakobi}\IEEEauthorrefmark{2}}
                     
   \IEEEauthorblockA{\IEEEauthorrefmark{1}%
                      International Institute of Information Technology, Hyderabad, India}
                      
\IEEEauthorblockA{\IEEEauthorrefmark{2}%
                     Department of Computer Science, 
                     Technion---Israel Institute of Technology, 
                     Haifa 3200003, Israel}\\
                     \texttt{shubhranshsinghvi2001@gmail.com}\\
 \thanks{The research was Funded by the European Union (ERC, DNAStorage, 101045114). Views and opinions expressed are however those of the authors only and do not necessarily reflect those of the European Union or the European Research Council Executive Agency. Neither the European Union nor the granting authority can be held responsible for them. \\
 
 A part of this work was presented at ITW 2022 \cite{SSBY22}
 }   
 }

\title{Conditional Entropies of\\ $k$-Deletion/Insertion Channels}
\date{\today}
\maketitle
\thispagestyle{empty}	
\pagestyle{empty}

\begin{abstract}
The channel output entropy of a transmitted  sequence is the entropy of the possible channel outputs and similarly the channel input entropy of a received sequence is the entropy of all possible transmitted sequences. The goal of this work is to study these entropy values for the $k$-deletion, $k$-insertion channels, where exactly $k$ symbols are deleted, inserted in the transmitted sequence, respectively. If all possible sequences are transmitted with the same probability then studying the input and output entropies is equivalent. For both the $1$-deletion and $1$-insertion channels, it is proved that among all sequences with a fixed number of runs, the input entropy is minimized for sequences with a skewed distribution of their run lengths and it is maximized for sequences with a balanced distribution of their run lengths. Among our results, we establish a conjecture by Atashpendar et al. which claims that for the $1$-deletion channel, the input entropy is maximized by the alternating sequences over all binary sequences. This conjecture is also verified for the $2$-deletion channel, where it is proved that constant sequences with a single run minimize the input entropy.
\end{abstract}

\maketitle

\section{Introduction}
In the last decade, channels that introduce insertion and deletion errors attracted significant attention due to their relevance to DNA storage systems~\cite{Anavy433524, GHP15, OAC17, PTH21, TWB20, YGM16}, where deletions and insertions are among the most dominant errors~\cite{HMG18, SOSAYY19}. The study of communication channels with insertion and deletion errors is also relevant to many other applications such as 
the synchronization of files and symbols of data streams~\cite{DolecekAnan_Sync_2007} and for cases of over-sampling and under-sampling at the receiver side~\cite{SSB16}. \emph{VT codes}, designed by Varshamov and Tenengolts~\cite{VT}, are the first family of codes that can correct a single deletion or a single insertion~\cite{L66}. Later, several works extended the scheme to correct multiple deletion errors, as well as substitutions and edit errors; see e.g.,~\cite{BGZ16, BGZ18, GS19, GGRW23, GH21, SWGY17, CJLW18, SB19, SRB21, SWWY23, SPCH22, SXG23 , RST23, S24}.  Under some applications, such as DNA storage systems, the problem of list decoding was studied as well, for example in~\cite{maria1, guruswami2020optimally, hayashi2018list, hanna2019list, liu2019list,  wachter2017list}. Under this setup, the decoder receives a channel output and returns a (short) list of possible codewords that includes the transmitted codeword. Even though many works considered channels with deletion errors and designed codes that correct such errors, finding optimal codes for these channels is yet far from being solved. 


Additionally, noisy channels that introduce deletion and insertion errors were also studied as part of \emph{the trace reconstruction problem}~\cite{BKK04} and the \emph{sequence reconstruction problem}~\cite{L011, L012}. In these problems, a codeword $\bfx$ is transmitted over the same channel multiple times. This transmission results in several noisy copies of $\bfx$, and the goal is to find the required minimum number of these noisy copies that enables the reconstruction of $\bfx$ with high probability or in the worst case. Theoretical bounds and other results for the trace reconstruction problem over the deletion channel were proved in several works such as~\cite{AVDG19, BKK04, D21, gabrys2018sequence, HPP18}, and other works also studied algorithms for the sequence reconstruction problem for channels that introduce deletion and insertion errors; see e.g.,~\cite{SB21, CDR21, CKNY21, yaakobi2013sequence}. Recently, \cite{SCKY24} studied the sequence reconstruction problem for Reed-Solomon codes over channels with substitution errors. 

This work studies several related problems to the capacity of the deletion and insertion channel and can contribute to their analysis. Assume a sequence $\bfx$ is transmitted over a communication channel and the channel output is $\bfy$. This work considers the \emph{output entropy}, which is the entropy of all possible channel outputs $\bfy$ for a given input sequence $\bfx$ and the \emph{input entropy}, first studied in~\cite{ABC18}, which is the entropy of all possible transmitted sequences $\bfx$, given a channel output $\bfy$. Note that unlike the output entropy, the input entropy also depends on the probability to transmit each sequence $\bfx$.  Hence, it is assumed in this work that this probability is equal for each sequence $\bfx$. Our main goal is to characterize these entropies, their expected values, and the sequences that maximize and minimize them for a combinatorial version of the insertion and the deletion channels, which are referred as the \emph{$k$-deletion channel} and the \emph{$k$-insertion channel}. In the $k$-deletion, $k$-insertion channel the number of deletions, insertions is exactly $k$ and the errors are equally distributed~\cite{BGSY21, G15}. It was shown in~\cite{ABC18} that the sequences that minimize the input entropy for the $1$-deletion and $2$-deletion channels, in the binary case, are the all-zeroes and all-ones sequences, under the equal transmission probability assumption.

The rest of the paper is organized as follows. In Section~\ref{sec:mot}, we motivate the task of studying the input and output entropies of the $k$-deletion and $k$-insertion channels by describing its relation to the problem of analysing the capacity of deletion channels. In Section~\ref{sec:pre}, we give the basic notations and definitions in the paper. In this section, we also give a formal definition of the problems studied in this paper. In Section~\ref{sec:charact}, we characterize the input entropies of the $k$-deletion and $k$-insertion channels and particularly for $k=1$~\&~$2$, we give a more elaborate characterization. Section~\ref{sec:extremum1} finds the extremum values and the sequences that attain these extremum values for the input entropy of both $1$-deletion and $1$-insertion channels over arbitrary alphabet size. Section~\ref{sec:extremum1} also studies the average input entropies for $k=1$. In Section~\ref{sec:extremum2}, we find the minimum values and the sequences that attain these minimum values for the input entropy of $2$-deletion channel and $k$-insertion channel over binary alphabet. Section~\ref{sec:conclusion} then concludes the paper with a brief summary of our contribution and discussion on future work and open problems.

\section{Motivation}\label{sec:mot}
The \emph{binary deletion channel} (BDC$_p$) models the setup where bits of a transmitted message are deleted from the message randomly and independently with probability $p$. Perhaps the most important question regarding the BDC is determining its capacity, i.e., the maximum achievable transmission rate over the channel that still allows recovering from the errors introduced by the channel, with high probability. In spite of many efforts (see the excellent surveys \cite{cheraghchi2020overview,mitzenmacher2009survey}), the capacity of the BDC$_p$ is still not known and it is an outstanding open challenge to determine it. 

In the extremal parameter regimes, the behavior of the capacity of the BDC is partially understood. When $p\rightarrow 0$ the capacity approaches $1-h(p)$~\cite{KMS10}. Mitzenmacher and Drinea~\cite{MD06} and Kirsch and Drinea~\cite{KD09} showed a method by which distributions on run lengths can be converted to codes for the BDC, yielding a lower bound of $\mathbb{C}(\mathrm{BDC}_p) > 0.1185 (1-p)$. 
Fertonani and Duman~\cite{FD10}, Dalai~\cite{D11} and Rahmati and Duman~\cite{RD15} used computer aided analyses based on the Blahut-Arimoto algorithm~\cite{A72,B72} to prove an upper bound of $\mathbb{C}(\mathrm{BDC}_p) < 0.4143 (1-p)$ in the high deletion probability regime ($p > 0.65$). Recently, Rubinstein and Con~\cite{RC23} showed that the Blahut-Arimoto algorithm can be implemented with a lower space complexity, allowing to extend the upper bound analyses, and proved an upper bound of $\mathbb{C}(\mathrm{BDC}_p) < 0.3745 (1-p)$ for all $p\geq 0.68$. Furthermore, they also showed that an extension of the Blahut-Arimoto algorithm can be used to select better run length distributions for Mitzenmacher and Drinea’s construction, yielding a lower bound of $\mathbb{C}(\mathrm{BDC}_p) > 0.1221(1-p)$.

In the recent work by Rubinstein and Con~\cite{RC23}, which improved the ones from~\cite{D11,FD10}, the main idea in deriving an upper bound on the capacity of the BDC$_p$ is to use some auxiliary channels. First, let $W_n^p$ be the binary deletion channel BDC$_p$ with fixed input length $n$ and let $\mathbb{C}_n(p)\triangleq \max I(X_1^n;Y)$ be its capacity where the maximum is taken over all input distributions on $n$ bits and $Y = W_n^p(X_1^n)$ is the channel output. It is known that for any $n$, $\mathbb{C}(\mathrm{BDC}_p) \leq \frac{\mathbb{C}_n(p)}{n}$~\cite{D11}. In the \emph{$k$-deletion channel}, denoted by $\kDel$, exactly $k$ symbols are deleted from the transmitted sequence. The $k$ symbols are selected uniformly at random out of the $\binom{n}{k}$ symbol positions. Denote the capacity of the $\kDel$ Channel by $\mathbb{C}^{\kDel}_{n} \triangleq \max I(X_1^n;Y)$, where $Y = \kDel(X_1^n)$ is the channel output. Fertonani and Duman~\cite{FD10} and Dalai~\cite{D11} proved the following inequality between the capacity values $\mathbb{C}_n(p)$ and $\mathbb{C}^{\kDel}_{n}$, which is given by \vspace{-1ex}
\begin{equation*}\label{eq:cap_bound}
\mathbb{C}_n(p) \leq \sum_{k=0}^n \binom{n}{k}p^{k}(1-p)^{n-k}\mathbb{C}^{\kDel}_{n}.
\end{equation*}
Since the channel $\kDel$ has fixed finite input and output sizes, it can be described by a finite transition matrix and thus its capacity can be calculated by the Blahut Arimoto algorithm~\cite{A72,B72}, and indeed the works~\cite{D11,FD10,RC23} used this property in order to bound the capacity values $\mathbb{C}^{\kDel}_{n}$ and thereby derive upper bounds on the capacity $\mathbb{C}(\mathrm{BDC}_p)$.

For a given input distribution $\mathsf{Pr}(\bfx)$ over $X_1^{n}$, the mutual information $I(X_1^n;Y)$, where $Y = \kDel(X_1^n)$, is given by
\begin{align*}
    I(X_1^n;Y) &= H(Y) - H(Y\vert X_1^n) \\
    &= H(Y) - \sum_{\bfx\in\{0,1\}^n} \mathsf{Pr}(\bfx)\, H(Y\vert X_1^n = \bfx),
\intertext{or, equivalently,}
    I(X_1^n;Y) &= H(X_1^n) - H(X_1^n \vert Y) \\
    &= H(X_1^n) - \sum_{\bfy\in\{0,1\}^{n-k}} \mathsf{Pr}(\bfy)\, H(X_1^n \vert Y = \bfy).
\end{align*}

Hence, one of the important tasks in studying the capacity of synchronization channels is to determine for every $\bfx\in\cX^n$ the conditional entropy $H(Y \vert X^n_1=\bfx)$ and similarly for every $\bfy\in\cY^*$ the conditional entropy $H(X^n_1 \vert Y = \bfy)$. Formally, for an input $\bfx\in \cX^n$, we refer to $H(Y \vert X^n_1=\bfx)$ as the \emph{output entropy} of the channel and for an output $\bfy\in \cY^*$, $H(X^n_1 \vert Y = \bfy)$ is the \emph{input entropy} of the channel. 

Note that as opposed to several discrete symmetric memoryless channels, such as the binary symmetric channel and the binary erasure channel, the input and output entropies for synchronization channels, and in particular for the insertion and deletion channels, depend on the specific choice of $\bfx$ and $\bfy$. Furthermore, while the output entropy depends solely on the channel $\cS$, the input entropy depends both on $\cS$ and the channel input distribution $\mathsf{Pr}(X^n_1)$. The case in which the channel input distribution is uniform is referred as \emph{uniform transmission}.

\begin{remark}
In this paper, we study the conditional entropies under uniform transmission. Nevertheless, a complete characterization of the channel capacity necessitates a deeper understanding of the optimal input distribution, which remains an important open problem and is left for future investigation.
\end{remark}

Among other applications, the output entropy can be used in list-decoders to rank the different codewords from the decoder's list by their transmission probability~\cite{guruswami2020optimally,wachter2017list}. The output entropy can also be used to assist with code design by prioritizing sequences with lower output entropy since they are likely to have higher successful decoding probability. Alternatively, when the goal is to encrypt the information, sequences with higher output entropies are preferred~\cite{ARR15}. In several of DNA reconstruction algorithms such as~\cite{BO21, SYSY20, TrellisBMA}, there is a limitation on the number of noisy copies that can be considered by the algorithm's decoder, due to design restrictions and run-time considerations. Therefore, in case the number of received noisy copies is greater than this limitation, a subset of these copies should be considered. Hence, to improve the accuracy of such algorithms, the input entropy of the channel outputs can be used to wisely select this subset of copies. Studying the input entropy was also investigated in~\cite{ABC18,ARR15} for analyzing the information leakage of a key by revealing any of its subsequences.

\section{Preliminaries and Problem Statement}\label{sec:pre}
Let $\mathbb{I}$ denote the indicator function. For a positive integer $n$, let $[n]\triangleq\{1,\ldots,n\}$ and let $\Sigma_q \triangleq \{0,1,\ldots,q-1\}$ be an alphabet of size $q$. For an integer $n\ge0$, let $\Sigma_q^n$ be the set of all words (sequences) of length $n$ over the alphabet $\Sigma_q$. For an integer $k$, $0\le k\le n$, a sequence $\bfy\in\Sigma_q^{n-k}$ is a  \emph{$k$-subsequence} of $\bfx\in\Sigma_q^n$ if $\bfy$ can be obtained by deleting $k$ symbols from $\bfx$. Similarly, a sequence $\bfy\in\Sigma_q^{n+k}$ is a  \emph{$k$-supersequence} of $\bfx\in\Sigma_q^n$ if $\bfx$ is a $k$-subsequence of~$\bfy$. Let $\bfx$ and $\bfy$ be two sequences of length $n$ and $m$ respectively such that $m < n$. The \emph{embedding number} of $\bfy$ in $\bfx$, denoted by $\omega_{\bfy}(\bfx)$, is defined as the number of distinct occurrences of $\bfy$ as a subsequence of $\bfx$. More formally, the embedding number is the number of distinct index sets, $(i_1, i_2, \ldots, i_{m})$, such that $1 \le i_1 < i_2 < \cdots < i_{m} \le n$ and $x_{i_1} = y_1, x_{i_2} = y_2, \dotsc, x_{i_{m}} = y_{m}$. For example, let $\bfx = \texttt{11220}$ and $\bfy = \texttt{120}$, then $\omega_{\bfy}(\bfx) = 4$. The \textit{$k$-insertion ball} centred at ${\bfx\in\Sigma_q^n}$, denoted by $I_k(\bfx)\subseteq \Sigma_q^{n+k}$, is the set of all $k$-supersequences of $\bfx$. Similarly, the \textit{$k$-deletion ball} centred at ${\bfx\in\Sigma_q^n}$, denoted by $D_k(\bfx)\subseteq \Sigma_q^{n-k}$, is the set of all $k$-subsequences of~$\bfx$.

Recall that in the \emph{$k$-deletion channel}, denoted by $\kDel$, exactly $k$ symbols are deleted from the transmitted sequence. The $k$ symbols are selected uniformly at random out of the $\binom{n}{k}$ symbol positions, where $n$ is the length of the transmitted sequence. Hence, the conditional probability of $\kDel$ Channel is

$$
\mathsf{Pr}_{\kDel}^{\mathsf{Out}}\{\bfy \vert \bfx\}=\frac{\omega_{\bfy}(\bfx)}{\binom{n}{k}},
$$ 
for all $\bfx\in\Sigma_q^n,\bfy\in\Sigma_q^{n-k}$. Similarly, for the \emph{$k$-insertion channel}, denoted by $\kIns$, exactly $k$ symbols are inserted to the transmitted sequence, while the locations and values of the $k$ symbols are selected uniformly at random out of the $\binom{n+k}{k}$ possible locations and $q^k$ possible options for the symbols. Therefore, the conditional probability of the $k$-insertion channel is

$$
\mathsf{Pr}_{\kIns}^{\mathsf{Out}}\{\bfy \vert \bfx\}=\frac{\omega_{\bfx}(\bfy)}{\binom{n+k}{k}q^k}, $$ %
for all $\bfx\in\Sigma_q^n,\bfy\in\Sigma_q^{n+k}$.

The goal of this work is to study the following values for the $k$-deletion and $k$-insertion channels.
\begin{problem}\label{prob1}
Find the following values for the channel $\kDel$:
\begin{enumerate}
    \item For all $\bfx\in\Sigma_q^n$, find its output entropy,
    $$\mathsf{H}^{\mathsf{Out}}_{\kDel} (\bfx) \triangleq H(Y \vert X^n_1=\bfx).$$
     
    \item Find the minimum, maximum, and average output entropy, 
    \begin{align*} \mathsf{min}^{\mathsf{Out}}_{\kDel}(n) &\triangleq \min_{\bfx\in\Sigma_q^n}\left\{\mathsf{H}^{\mathsf{Out}}_{\kDel} (\bfx)\right\}, \\
    \mathsf{max}^{\mathsf{Out}}_{\kDel}(n) &\triangleq \max_{\bfx\in\Sigma_q^n}\left\{\mathsf{H}^{\mathsf{Out}}_{\kDel} (\bfx)\right\}, \\
    \mathsf{avg}^{\mathsf{Out}}_{\kDel}(n) &\triangleq \mathbb{E}_{\bfx\in\Sigma_q^n}\left\{\mathsf{H}^{\mathsf{Out}}_{\kDel} (\bfx)\right\}.
    \end{align*}
     
    \item For all $\bfy\in\Sigma_q^{n-k}$, find its input entropy with input distribution $\mathsf{Pr}(X^n_1)$,
    $$\mathsf{H}^{\mathsf{In}}_{\kDel,\mathsf{Pr}(X^n_1)} (\bfy) \triangleq H(X^n_1 \vert Y = \bfy).$$
     
    \item Find the minimum, maximum, and average input entropy with input distribution $\mathsf{Pr}(X^n_1)$,  
    \begin{align*}
    \mathsf{min}^{\mathsf{In}}_{\kDel}\left(n,\mathsf{Pr}\left(X^n_1\right)\right) &\triangleq \min_{\bfy\in\Sigma_q^{n-k}}\left\{\mathsf{H}^{\mathsf{In}}_{\kDel,\mathsf{Pr}(X^n_1)} (\bfy)\right\}, \\
    \mathsf{max}^{\mathsf{In}}_{\kDel}\left(n,\mathsf{Pr}\left(X^n_1\right)\right) &\triangleq \max_{\bfy\in\Sigma_q^{n-k}}\left\{\mathsf{H}^{\mathsf{In}}_{\kDel,\mathsf{Pr}(X^n_1)} (\bfy)\right\},\\
     \mathsf{avg}^{\mathsf{In}}_{\kDel}\left(n,\mathsf{Pr}\left(X^n_1\right)\right) &\triangleq \mathbb{E}_{\bfy\in\Sigma_q^{n-k}}\left \{\mathsf{H}^{\mathsf{In}}_{\kDel,\mathsf{Pr}(X^n_1)} (\bfy)\right \}. 
    \end{align*}
      
\end{enumerate}
The equivalent values and notations are defined similarly for the $k$-insertion channel, $\kIns$.
\end{problem}

In the following lemma, we show the relation between the input entropy and the output entropy of the $k$-insertion and the $k$-deletion channels under uniform transmission. 
\begin{lemma}\label{lem:uni_equiv}
For uniform transmission over the $k$-insertion, {$k$-deletion} channel and for any channel output $\bfy$, it holds that 
$$\mathsf{H}^{\mathsf{In}}_{\kDel,\mathsf{Pr}(X^n_1)} (\bfy) = \mathsf{H}^{\mathsf{Out}}_{\kIns} (\bfy)~\textrm{ and }~\mathsf{H}^{\mathsf{In}}_{\kIns,\mathsf{Pr}(X^n_1)} (\bfy) = \mathsf{H}^{\mathsf{Out}}_{\kDel} (\bfy). $$ 
\end{lemma}

\begin{IEEEproof}
Since any sequence $\bfx$ of length $n$ can be transmitted with the same probability, it can be shown that any channel output $\bfy$ of length $n-k$, $n+k$ can be obtained with the same probability as an output of the channel $\kDel$, $\kIns$, respectively. Thus, the results can be derived directly from the definitions.
\end{IEEEproof}

\begin{remark}
Building on Lemma~\ref{lem:uni_equiv}, the remainder of the paper analyzes the input entropy of the $k$-insertion and $k$-deletion channels under the uniform input distribution. Accordingly, we omit the $\mathsf{Pr}(X_1^n)$ term from the notation for input entropy throughout the rest of the paper.
\end{remark}

\subsection{Related Work}
Atashpendar et al.~\cite{ABC18} studied the minimum input entropy of the $\kDel$ channel under uniform transmission for the case where $k \in \{ 1, 2\}$ and $q=2$. In their work, they presented an efficient algorithm to count the number of occurrences of a sequence $\bfy$ in any given supersequence $\bfx$. They also provided an algorithm that receives a sequence $\bfy$, computes the set of all of its supersequences of specific length, characterizes the distribution of their embedding numbers and clusters them by their Hamming weight. Lastly, they proved that the all-zero and the all-one sequences minimize the input entropy of the $1\textrm{-}\mathsf{Del}$ and the $2\textrm{-}\mathsf{Del}$ channels.

In another related line of work, Atashpendar et al.~\cite{AMR18} studied the minimum input entropy of the $\kDel$ channel under uniform transmission within an asymptotic regime where the output length is fixed while the input length and the parameter $k$ tend to infinity. Their approach employs analytic-combinatorial techniques based on the hidden word statistics framework developed by Flajolet et al.\cite{FSV06}. In particular, they analyze the moments of the posterior distribution (i.e., the conditional probability of an input sequence given a fixed output sequence) and demonstrate that these are governed by an input-dependent autocorrelation measure. They show that the all-zero and all-one sequences maximize this autocorrelation measure, which in turn leads to the minimization of the input entropy.

\subsection{Our Contributions}

This paper makes the following key contributions:

\begin{enumerate}
\item \textbf{General entropy characterizations.} We derive explicit expressions for the input and output conditional entropies of the $k$-deletion and $k$-insertion channels under the assumption of uniform input distribution, expressing them in terms of embedding numbers of sequences.

\item \textbf{Closed-form formulas for $k=1,2$.}
\begin{itemize}
\item For the 1-deletion and 1-insertion channels, we obtain entropy formulas invariant under permutations of the run-length profile, reducing the analysis to combinatorial properties of runs.
\item For the 2-deletion channel, we  extend prior work of Atashpendar et al.~\cite{ABC18} by providing new closed-form entropy expressions that incorporate the structure of alternating segments.
\end{itemize}

\item \textbf{Extremal entropy values and extremizing sequences.}
\begin{itemize}
\item We fully characterize the sequences that attain the minimum and maximum input entropy in both $1$-deletion and $1$-insertion channels: single-run sequences minimize input entropy, while  alternating sequences maximize it. In particular, we resolve the conjecture of Atashpendar \emph{et al.} that alternating sequences maximize the  input entropy for $1$-deletion channel.
\item We prove that single-run sequences uniquely minimize the input entropy for $2$-deletion channel and determine the exact minimum value. In doing so, we improve upon the previous characterization in the literature, employ a refined alternating-segment analysis, and rigorously establish both the minimizing sequences and the exact minimum input entropy value for the $2$-deletion channel.
\end{itemize}

\item \textbf{Average entropy} We derive exact formulas for the average input entropy of $1$-insertion and $1$-deletion channels by enumerating the contribution of each possible run length across all sequences, and we establish closed-form lower bounds.

\item \textbf{Results for higher-order channels.} For arbitrary $k\geq2$, we prove that constant sequences uniquely minimize the input entropy for the $k$-insertion channel. 
\end{enumerate}

\section{Characterization of the Input Entropy}\label{sec:charact}
This section first presents a complete characterization of the input entropy of the channels $\kDel$, $\kIns$ and gives an explicit expression of these entropies for any channel output $\bfy$. We then further analyse the entropy expressions for $k=1,2$. Let $\cS \subseteq  \Sigma_q^n$ and define $\mathsf{W}_{\cS}(\bfy) \triangleq \sum_{\bfx \in \cS } \omega_{\bfy}(\bfx) \cdot \log \left( {\omega_{\bfy}(\bfx)}\right)$. 

\begin{lemma} \label{lm:entropyIn-k-del}
For any integers $n$ and $k$, such that $k \le n$, and for any sequence $\bfy \in \Sigma_q^{n-k}$, we have that
$$\mathsf{H}^{\mathsf{In}}_{\kDel} (\bfy) 
 = 
\log \left( \binom{n}{k} q^k\right) -\frac{\mathsf{W}_{I_k(\bfy)}(\bfy)}{\binom{n}{k}q^k}.$$
\end{lemma}
\begin{IEEEproof}
\begin{align*}
\mathsf{H}^{\mathsf{In}}_{\kDel} (\bfy) &=
- \sum_{\bfx \in I_k(\bfy) }\left( \mathsf{Pr}_{\kDel}^{\mathsf{In}}\{\bfx \vert \bfy\} \cdot \log \left( \mathsf{Pr}_{\kDel}^{\mathsf{In}}\{\bfx \vert \bfy\} \right) \right) 
\\&=
- \sum_{\bfx \in I_k(\bfy) } \frac{\omega_{\bfy}(\bfx)}{\binom{n}{k} q^k } \cdot \log \left( \frac{\omega_{\bfy}(\bfx)}{\binom{n}{k} q^k }\right)  
\\ & = 
\log \left( \binom{n}{k} q^k\right)  -\frac{1}{\binom{n}{k}q^k}\sum_{\bfx \in I_k(\bfy) } \omega_{\bfy}(\bfx) \cdot \log \left( {\omega_{\bfy}(\bfx)}\right).
\end{align*} 
\end{IEEEproof}
Similarly, we have the next lemma for the $\kIns$ Channel. 
\begin{lemma}
\label{lm:entropyIn-k-ins}
For any integers $n$ and $k$, such that $k \le n$, and for any sequence $\bfy \in \Sigma_q^{n+k}$, we have that
$$\mathsf{H}^{\mathsf{In}}_{\kIns} (\bfy) = \log  \binom{n+k}{k} -\frac{\mathsf{W}_{D_k(\bfy)}(\bfy)}{\binom{n+k}{k}}.$$
\end{lemma}
\begin{IEEEproof}
\begin{align*}
\mathsf{H}^{\mathsf{In}}_{\kIns} (\bfy) &=
- \sum_{\bfx \in D_k(\bfy) }\left( \mathsf{Pr}_{\kIns}^{\mathsf{In}}\{\bfx \vert \bfy\} \cdot \log \left( \mathsf{Pr}_{\kIns}^{\mathsf{In}}\{\bfx \vert \bfy\} \right) \right) 
\\&=
- \sum_{\bfx \in D_k(\bfy) } \frac{\omega_{\bfx}(\bfy)}{\binom{n+k}{k} } \cdot \log \left( \frac{\omega_{\bfx}(\bfy)}{\binom{n+k}{k}}\right)  
\\ & = 
\log \left(\binom{n+k}{k}\right)  -\frac{1}{\binom{n+k}{k}}\sum_{\bfx \in D_k(\bfy) } \omega_{\bfx}(\bfy) \cdot \log \left( {\omega_{\bfx}(\bfy)}\right).
\end{align*} 
\end{IEEEproof}

The next claim follows from Lemma \ref{lm:entropyIn-k-del} and \ref{lm:entropyIn-k-ins}. 
\begin{claim} \label{claim:HW}
    For $k \leq n$, we have that 
\begin{align*}
\underset{\bfy \in \Sigma_q^{m}}{\argmin}~\mathsf{H}^{\mathsf{In}}_{\kDel} (\bfy) &= \underset{\bfy \in \Sigma_q^{m}}{\argmax}~\mathsf{W}_{I_k(\bfy)}(\bfy), \\ 
\underset{\bfy \in \Sigma_q^{m}}{\argmax}~\mathsf{H}^{\mathsf{In}}_{\kDel} (\bfy) &= \underset{\bfy \in \Sigma_q^{m}}{\argmin}~\mathsf{W}_{I_k(\bfy)}(\bfy),\\
\underset{\bfy \in \Sigma_q^{m}}{\argmin}~\mathsf{H}^{\mathsf{In}}_{\kIns} (\bfy) &= \underset{\bfy \in \Sigma_q^{m}}{\argmax}~\mathsf{W}_{D_k(\bfy)}(\bfy), \\ 
\underset{\bfy \in \Sigma_q^{m}}{\argmax}~\mathsf{H}^{\mathsf{In}}_{\kIns} (\bfy) &= \underset{\bfy \in \Sigma_q^{m}}{\argmin}~\mathsf{W}_{D_k(\bfy)}(\bfy).
\end{align*}
\end{claim}

For a sequence $\bfx$, a \emph{run} of $\bfx$ is a maximal subsequence of identical consecutive symbols within $\bfx$. The number of runs in $\bfx$ is denoted by $\rho(\bfx)$.
We denote by $\Sigma_{q,R}^n$, the set of sequences $\bfx \in \Sigma_q^n$, such that $\rho(\bfx)=R$. It is well known that the number of such sequences is $ \vert \Sigma_{q,R}^n \vert  = {\binom{n-1}{R-1}} q (q-1)^{R-1}$. For a sequence $\bfx \in \Sigma_{q,R}^n$, its \emph{run length profile}, denoted by $\cR\cL(\bfx)$, is the vector of the lengths of the runs in $\bfx$. That is, 
$
\cR\cL(\bfx) \triangleq  (r_1,r_2,\ldots, r_R),
$
where $r_{i}$, for $i\in[R]$, denotes the length of the $i$-th run of $\bfx$. For example, let $q=4$ and $\bfx = \texttt{311221110}$, then $\cR\cL(\bfx) = (1, 2, 2, 3, 1)$.

For $k=1$, we have the next two theorems for the $1\textrm{-}\mathsf{Del}$ and $1\textrm{-}\mathsf{Ins}$ channels.

\begin{theorem}
\label{thm:entropy-1-deletion}
Let $\bfy\in\Sigma_{q}^{n-1}$ and $\cR\cL(\bfy) = (r_1, r_2, \ldots, r_{\rho(\bfy)})$. Then,  $\mathsf{H}^{\mathsf{In}}_{1\textrm{-}\mathsf{Del}} (\bfy)$ is invariant to permutations of $\cR\cL(\bfy)$ and it is given by the following expression
\begin{align*}
\mathsf{H}^{\mathsf{In}}_{1\textrm{-}\mathsf{Del}} (\bfy) = \log(nq) - \frac{1}{nq}\sum_{i=1}^{\rho(\bfy)}(r_{i}+1)\log(r_{i}+1).
\end{align*}
\end{theorem}
\begin{IEEEproof}
From~\cite{BGSY21, SBGYY22}, it is known that $I_1(\bfy)$ contains $\rho(\bfy)$ sequences that can be obtained by prolonging an existing run, each with embedding number $r_{i}+1$ for $1 \le i \le \rho(\bfy)$. The embedding number of the remaining $ \vert I_{1}(\bfy) \vert  - \rho(\bfy) =  q+ \vert \bfy \vert (q-1) - \rho(\bfy)$ sequences is one. Hence, upon substituting these embedding numbers in the expression from Lemma~\ref{lm:entropyIn-k-del}, we get 
\begin{align*}
\mathsf{H}^{\mathsf{In}}_{1\textrm{-}\mathsf{Del}} (\bfy) &= \log \left( nq\right)  -\frac{1}{nq}\sum_{\bfx \in I_1(\bfy) } \omega_{\bfy}(\bfx) \cdot \log \left( {\omega_{\bfy}(\bfx)}\right)\\
&= \log(nq) - \frac{1}{nq}\left(\sum_{i=1}^{\rho(\bfy)}(r_{i}+1)\log(r_{i}+1)\right).
\end{align*}
Therefore, as the entropy expression does not depend on the order of the runs, it is invariant to permutations of $\cR\cL(\bfy)$.  
\end{IEEEproof}

\begin{theorem}
\label{thm:entropy-1-insertion}
Let $\bfy\in\Sigma_{q}^{n+1}$ and $\cR\cL(\bfy) = (r_1, r_2, \ldots, r_{\rho(\bfy)})$. Then, $\mathsf{H}^{\mathsf{In}}_{1\textrm{-}\mathsf{Ins}} (\bfy)$ is invariant to permutations of $\cR\cL(\bfy)$ and it is given by the following expression
\begin{align*}
\mathsf{H}^{\mathsf{In}}_{1\textrm{-}\mathsf{Ins}} (\bfy) = \log(n+1) -\frac{1}{n+1}\sum_{i=1}^{\rho(\bfy)}r_{i}\log(r_{i}).
\end{align*}
\end{theorem}

\begin{IEEEproof}
From~\cite{BGSY21, SBGYY22}, it is known that $D_1(\bfy)$ contains $\rho(\bfy)$ sequences with embedding numbers, $r_{1}, r_{2}, \ldots, r_{\rho(\bfy)}$, obtained by shortening a run. Hence, upon substituting these embedding numbers in the expression from Lemma~\ref{lm:entropyIn-k-ins}, we get 
\begin{align*}
\mathsf{H}^{\mathsf{In}}_{1\textrm{-}\mathsf{Ins}} (\bfy) &= \log(n+1) -\frac{1}{n+1}\sum_{\bfx \in D_1(\bfy) } \omega_{\bfx}(\bfy) \cdot \log \left( {\omega_{\bfx}(\bfy)}\right)\\
&= \log(n+1) -\frac{1}{n+1}\left(\sum_{i=1}^{\rho(\bfy)}(r_{i})\log(r_{i})\right).
\end{align*}
Therefore, as the entropy expression does not depend on the order of the runs, it is invariant to permutations of $\cR\cL(\bfy)$.  
\end{IEEEproof}
A subsequence $\bfx[i,j]$ is said to be an \textit{alternating segment} \cite{DTE21, DTE22} if $\bfx[i,j]$ is a sequence of alternating distinct symbols $a, b \in \Sigma_q$. Note that $\bfx[i,j]$ is a \textit{maximal alternating segment} if $\bfx[i,j]$ is an alternating segment and $\bfx[i-1,j]$, $\bfx[i,j+1]$ are not. Let the number of maximal alternating segments of $\bfx$ be denoted by $\alpha(\bfx)$. For example, in $\bfx= \texttt{00110100}, \rho(\bfx) = 5 $ and $\alpha(\bfx) = 4$, where the four maximal alternating segments are $\texttt{0,01,1010,0}$. Note that for $\bfx \in \Sigma_2^n$, $\alpha(\bfx) +\rho(\bfx) = n + 1$ \cite{DTE21, DTE22}. Let $\cA\cL(\bfx) \triangleq (a_1,a_2, \ldots, a_{\alpha(\bfx)})$ be the vector denoting the lengths of maximal alternating segments of $\bfx$. 

Let $\bfy\in\Sigma_{2}^{m}$ with $\cA\cL(\bfy)=(a_1,a_2,\ldots,a_A)$. For $i\in[1,A-1]$, let $f_i$ denote the smallest index in $[i+1,A]$ such that $a_{_i}>1$, and let $f_A=A$. If such an index does not exist then $f_i=A$. Similarly, for $i \in [2,A]$, let $b_i$ denote the largest index in $[1,i-1]$ such that $a_{b_i}>1$ and let $b_1=1$. If such an index does not exist then $b_i=1.$  For example, let $\bfy = \texttt{000101011}$, then the indices $\{f_i\}_{i=1}^{4}$ and $\{b_i\}_{i=1}^{4}$ along with the maximal alternating segments are shown in the following table:
\vspace{0.35cm}
\begin{center}
\begin{tabular}{|c|c|c|c|c|}
\hline
$i$ & $1$ & $2$ & $3$ & $4$ \\
\hline
$a_i$ & $1$ & $1$ & $6$ & $1$ \\
\hline
$f_i$  & $3$ & $3$ & $4$ & $4$ \\
\hline
$b_i$  & $1$ & $1$ & $1$ & $3$ \\
\hline
\end{tabular}
\end{center}

\begin{remark}
\label{bifi}
    Note that $b_i = i$ if and only if $i = 1$, and similarly  $f_i = i$ if and only if $i = A$.
\end{remark}

\begin{lemma}
\label{correction}
Let $\bfy\in\Sigma_{2}^{n-2}$, $\cA\cL(\bfy)=(a_1,a_2,\ldots,a_A)$ and $\bfx \in I_2(\bfy)$. If $\cA\cL(\bfx)=(a_1,\ldots,a_i+2,a_{i+1},\ldots,a_A)$, i.e., two additional runs of length one are inserted to $\bfy$ within an existing alternating segment, then
\begin{align*}
\omega_{\bfy}(\bfx) = 1+ \sum_{j = b_i}^{f_i} a_j -(a_{b_i}-1)\cdot\{\mathbb{I}_{b_i\neq i}\}-(a_{f_i}-1)\cdot\{\mathbb{I}_{f_i\neq i}\}.
\end{align*}
\end{lemma}
\begin{IEEEproof}
Let $\cR\cL(\bfy)=(r_1,\ldots,r_R)$. When $a_i= 1$, the two new runs are inserted into an existing run. Thus, it can be readily verified that $\omega_{\bfy}(\bfx) = 1+ \sum_{j = b_i}^{f_i} a_j - (a_{f_i}-1) - (a_{b_i}-1) = 3 + \sum_{j = b_i+1}^{f_i-1} a_j$. Observe that if $b_i = i$ then $a_{b_i} - 1 = a_i - 1 = 0$, and similarly if $f_i = i$ then $a_{f_i} - 1 = a_i - 1 = 0$. 

For $a_i>1$, if the two new runs are inserted after the $t$-th run then $\cR\cL(\bfx)=(r_1,\ldots,r_t,1,1,r_{t+1},\ldots r_R)$. In \cite{ABC18}, ${\omega_{\bfy}(\bfx)}$ was characterized as $r_{t}+r_{t+1}+1$. In the case where $r_t=1$ or  $r_{t+1}=1$, the analysis is a bit harder and the characterization is slightly different as follows. Let $\bfz$ denote the subsequence of $\bfy$ from the $(b_i)$-th maximal alternating segment to the $(f_i)$-th maximal alternating segment. Since $\bfx$ is obtained from $\bfy$ by adding two runs of length one after the $t$-th run, we can consider the subsequence $\bfz'$ of $\bfx$ which is the sequence $\bfz$ with these additional two runs.  It can be verified that deleting any two bits in $\bfx$ that do not belong to $\bfz'$  will result with a sequence which is different than $\bfy$. Therefore, $\omega_{\bfy}(\bfx) = \omega_{\bfz}(\bfz').$ It can be readily verified that $\omega_{\bfz}(\bfz') = 1+ \sum_{j = b_i}^{f_i} a_j -(a_{b_i}-1)\cdot\{\mathbb{I}_{i\neq 1}\}-(a_{f_i}-1)\cdot\{\mathbb{I}_{i\neq A}\}$. Hence, from Remark \ref{bifi}, the result follows. 
\end{IEEEproof}

For $k=2$, we have the next two theorems for the $2\textrm{-}\mathsf{Del}$ and $2\textrm{-}\mathsf{Ins}$ channels. 

\begin{theorem} \label{thm:entropy-2-deletion}
Let $\bfy\in\Sigma_{2}^{n-2}$, $\cR\cL(\bfy) = (r_1, r_2, \ldots, r_{R})$, and $\cA\cL(\bfy)=(a_1,a_2,\ldots,a_{A})$. Then, $\mathsf{H}^{\mathsf{In}}_{2\textrm{-}\mathsf{Del}} (\bfy)$ is given by the following expression
\begin{align*}
\mathsf{H}^{\mathsf{In}}_{2\textrm{-}\mathsf{Del}} (\bfy) &= \log \left(2n(n-1)\right)  -\frac{\mathsf{W}_{I_2(\bfy)}(\bfy)}{2n(n-1)},
\end{align*}
where 
\begin{align*}
\mathsf{W}_{I_2(\bfy)}(\bfy) &= \sum_{i=1}^{R}\cF\left(\binom{r_i+2}{2}\right) + \sum_{(i,j)\in\{(i,j):1 \leq i < j\leq R\}} \cF\left((r_i+1)(r_j+1)\right)\\
&+ \sum_{i=1}^{R}(n+1-R-r_i)\cF(r_i+1)\\
&+ \sum_{i=1}^{A} \cF\left(1+ \sum_{j = b_i}^{f_i} a_j -(a_{b_i}-1)\cdot\{\mathbb{I}_{{b_i}\neq i}\}-(a_{f_i}-1)\cdot\{\mathbb{I}_{{f_i}\neq i}\}\right),
\end{align*}
and $\cF(a) = a\log(a)$.
\end{theorem}

\begin{IEEEproof}
    From Lemma \ref{correction} and \cite{ABC18}, we have the following cases: 
    
    \begin{enumerate}
        \item For all $i \in [R]$, we have one supersequence with embedding number $\binom{r_i+2}{2}$ obtained by prolonging the $i$-th run by length $2$.  
        \item For all $1 \leq i < j\leq R$, we have one supersequence with embedding number $(r_i + 1)(r_j + 1)$ obtained by prolonging the $i$-th and $j$-th run by length $1$.         
        \item For all $i \in [R]$, we have $1+(n-2)-(r_i-1)-(R-1) = n+1-R-r_i$ supersequences with embedding number $r_i+1$ obtained by prolonging the $i$-th run by length $1$ and splitting some different run or by creating a new run at the first or last position.  
        \item The number of distinct supersequences obtained by prolonging and splitting the same run, i.e., by extending an alternating segment by length 2 is $A = n-1-R$. The embedding number for these supersequences is given by Lemma \ref{correction}.
        \item Let $\Tilde{r}_i = r_i - 1 + \mathbb{I}_{i=1} + \mathbb{I}_{i=R}$. We have $\underset{1\leq i<j\leq R}{\sum}\Tilde{r}_i\Tilde{r}_j + \sum_{i=1}^{R}\binom{\Tilde{r}_i+1}{2}$ supersequences with embedding number $1$ obtained by splitting two runs or by splitting the same run twice. For $R = 1$, in \cite{ABC18}, the number of such supersequences was  mischaracterized as $\binom{r_1+1}{2} < \binom{\Tilde{r}_1+1}{2}$. 
    \end{enumerate}
The result then follows by substituting these embedding numbers and their respective counts into the input entropy expression derived in Lemma~\ref{lm:entropyIn-k-del}. 
\end{IEEEproof}

\begin{theorem}
\label{thm:entropy-double-insertion}
Let $\bfy\in\Sigma_{2}^{n+2}$, $\cR\cL(\bfy) = (r_1, r_2, \ldots, r_{R})$, and $\cA\cL(\bfy)=(a_1,a_2,\ldots,a_{A})$. Then, $\mathsf{H}^{\mathsf{In}}_{2\textrm{-}\mathsf{Ins}} (\bfy)$ is given by the following expression
\begin{align*}
\mathsf{H}^{\mathsf{In}}_{2\textrm{-}\mathsf{Ins}} (\bfy) = \log  \binom{n+2}{2} -\frac{\mathsf{W}_{D_2(\bfy)}(\bfy)}{\binom{n+2}{2}}, 
\end{align*}
where 
\begin{align*}    
\mathsf{W}_{D_2(\bfy)} &= \sum_{i=1}^{R} \cF\left(\binom{r_i}{2} \right)\cdot\mathbb{I}_{r_i\geq 2} + \sum_{(i,j)\in\{(i,j):1 \leq i < j-1\leq R\}} \cF\left(r_ir_j\right)\\
+& \sum_{i=1}^{A} \cF\left(\sum_{j = b_i}^{f_i} a_j -(a_{b_i}-1)\cdot\{\mathbb{I}_{{b_i}\neq i}\}-(a_{f_i}-1)\cdot\{\mathbb{I}_{{f_i}\neq i}\}-1\right) \cdot \mathbb{I}_{a_i>2}\\
&+ \sum_{i=1}^{A} \cF((i+b_i-1)(f_i-i+1))\cdot \mathbb{I}_{a_i = 2},
\end{align*}
and $\cF(a) = a\log(a)$.
\end{theorem}

\begin{IEEEproof}
 Note that by deleting two bits from a single run or two non adjacent runs or two adjacent runs we can generate all possible subsequences of any sequence. Now consider the following cases: 
     

    \begin{enumerate}
        \item For $i \in [R]$, if $r_i\geq 2$ then we have one subsequence with embedding number $\binom{r_i}{2}$ obtained by shortening the $i$-th run by $2$.
        \item For all $1 \leq i < j-1\leq R$, we have one subsequence with embedding number $r_ir_j$ obtained by shortening the non-adjacent $i$-th and $j$-th runs by length $1$.  
        \item  For $i \in [A]$, if $a_i > 2$ then from Lemma \ref{correction}, we have one subsequence with embedding number  $\sum_{j = b_i}^{f_i} a_j -(a_{b_i}-1)\cdot\{\mathbb{I}_{{b_i}\neq i}\}-(a_{f_i}-1)\cdot\{\mathbb{I}_{{f_i}\neq i}\}-1$ obtained by shortening the $i$-th maximal alternating segment by length $2$. If $a_i = 2$ then we have one subsequence with embedding number $(i-b_i+1)(f_i-i+1)$ obtained by shortening each of the adjacent runs that compose the $i$-th alternating segment by length one.
    \end{enumerate}
The result then follows by substituting these embedding numbers and their respective counts into the input entropy expression derived in Lemma~\ref{lm:entropyIn-k-ins}. 
\end{IEEEproof}

\section{Extremum and Average Values for $k=1$}\label{sec:extremum1}
In this section, we find the channel outputs that have maximum and minimum input entropy among all sequences in $\Sigma_{q,R}^m$, where $R\in[m]$, for $m \in \{ n-1, n+1 \}$. Then as corollaries, we find the maximum and minimum values of the input entropies of the channels $1\textrm{-}\mathsf{Del}$ and $1\textrm{-}\mathsf{Ins}$, and the sequences that attain these entropies. We first introduce the notion of balanced and skewed sequences. It is said that $\bfx \in \Sigma_{q,R}^n$ is \textit{skewed} if it consists of $R-1$ runs of length one and a single run of length $n-(R-1)$; $\bfx$ is \textit{balanced} if it consists of $r\equiv n \bmod R$ runs of length $\ceil{\frac{n}{R}}$ and the remaining $R - r$ runs are of length $\floor{\frac{n}{R}}$.

\begin{lemma}\label{lm:min-sum-R-runs}
Let $a\in \mathbb{Z}_{+}$. For $R\in [m]$ and $r\equiv m \bmod R$,
 \begin{align*}
\begin{split}
\min_{\bfy \in \Sigma_{q,R}^{m}}\sum_{i=1}^{R}(r_i+a) &= r\left(\ceil{\frac{m}{R}}+a\right)\log\left(\ceil{\frac{m}{R}}+a\right) \\
&+(R-r)\left(\floor{\frac{m}{R}}+a\right)\log\left(\floor{\frac{m}{R}}+a\right),
\end{split}
\end{align*}
 where $\cR\cL(\bfy) = (r_1,r_2,\ldots, r_R)$.
\end{lemma}
\begin{IEEEproof} 

Let $\bfy \in \Sigma_{q,R}^{m}$ be a sequence which minimizes the summation and assume its run length profile is $(r_1,r_2,\ldots,r_R)$. Assume to the contrary that $\bfy$ is not one of the balanced sequences. Then, there exist indices $\ell\neq s$ such that $ r_s< \floor{\frac{m}{\rho(\bfy)}} $ and $\ceil{\frac{m}{\rho(\bfy)}} < r_\ell$ and in particular $r_\ell - r_s \geq2$. Consider a sequence $\bfy' \in \Sigma_{q,R}^{m}$, with run length profile $(r_1',r_2',\ldots,r_R')$, where $r'_\ell = r_\ell-1$, $r'_s=r_s+1$, and for any $k \notin \{ \ell, s \}$, $r'_k=r_k$. Next, consider the difference, $\sum_{i=1}^{R}(r_i'+a) - \sum_{i=1}^{R}(r_i+a)$, which is equal to 
\begin{align*}
& \big(r_{\ell}\log r_{\ell}+ (r_{s}+2)\log(r_{s}+2)  -(r_{\ell}  +  1)\log(r_{\ell}  +  1)  -   (r_{s}  +  1)\log(r_{s}  +  1) \big).
\end{align*}
Let $g(r) \triangleq (r+1)\log(r+1) - r\log(r)$, and note that $g$ is an increasing function w.r.t. $r$. In addition, since $r_\ell > r_s+1$,
\begin{align*}
\sum_{i=1}^{R}(r_i'+a) - \sum_{i=1}^{R}(r_i+a)  =  g(r_s+1) - g(r_\ell) < 0.
\end{align*}
 
This is a contradiction to the assumption on $\bfy$. Therefore, among all sequences in $\Sigma_{q,R}^{m}$, $\sum_{i=1}^{R}(r_i+a)$ is minimized by balanced sequences, and the value can be simply derived from the run length profile of the balanced channel outputs. 
\end{IEEEproof}

\begin{lemma}\label{lm:max-sum-R-runs}
Let $a\in \mathbb{Z}_{+}$. For $R\in [m]$ and $r\equiv m \bmod R$,
 \begin{align*}
   \max_{\bfy \in \Sigma_{q,R}^{m}}\sum_{i=1}^{R}(r_i+a) &= (m-R+1+a)\log(m-R+1+a)\\
   &+ (a+1)\log(a+1)(R-1),
\end{align*}
 where $\cR\cL(\bfy) = (r_1,r_2,\ldots, r_R)$.
\end{lemma}
\begin{IEEEproof}
Let $\bfy \in \Sigma_{q,R}^{m}$ be a sequence with maximal sum and assume its run length profile is $(r_1,r_2,\ldots,r_R)$. Assume to the contrary that $\bfy$ is not one of the skewed sequences. Then there exist indices $\ell\neq s$ such that $1< r_s \leq r_\ell < m-(R-1)$. Consider the sequence $\bfy' \in \Sigma_{q,R}^{m}$, with run length profile $(r_1',r_2',\ldots,r_R')$, where
\begin{align*}
r_i' = \begin{cases} 
      r_i+1 & i = \ell \\
      r_i-1 & i = s\\
      r_i & \textrm{otherwise.} 
 \end{cases}
 \end{align*}
Next, consider the difference, $\sum_{i=1}^{R}(r_i'+a) - \sum_{i=1}^{R}(r_i+a)$ which is equal to
\begin{align*}
& \big(r_{s}\log r_{s}+ (r_{\ell}+2)\log(r_{\ell}+2)  -(r_{s}  +  1)\log(r_{s}  +  1)  -   (r_{\ell}  +  1)\log(r_{\ell}  +  1) \big).
\end{align*}
Let $g(r) \triangleq (r+1)\log(r+1) - r\log(r)$, and note that $g$ is an increasing function w.r.t. $r$. In addition, since $r_\ell + 1 > r_s$,
\begin{align*}
\sum_{i=1}^{R}(r_i'+a) - \sum_{i=1}^{R}(r_i+a)  = g(r_\ell+1) - g(r_s) > 0.
\end{align*}
This is a contradiction to the assumption on $\bfy$. Therefore, among all sequences in $\Sigma_{q,R}^{m}$, $\sum_{i=1}^{R}(r_i+a)$ is maximized by skewed sequences, and the value can be simply derived from the run length profile of the skewed channel outputs. 
\end{IEEEproof}

\subsection{Extremum Values for the $1$-Deletion Channel}
\subsubsection{Maximum Input Entropy} This subsection studies the channel outputs $\bfy \in \Sigma_q^m$ that maximize the input entropy of the $1\textrm{-}\mathsf{Del}$ channel, where $m\triangleq n-1$. 

The next lemma simply follows from Claim~\ref{claim:HW} and by substituting $a = 1$ in Lemma~\ref{lm:min-sum-R-runs} and thus we omit the proof. 

\begin{lemma}\label{lm:max-entropy-R-runs}
For $R\in [m]$ and $r\equiv m \bmod R$, the maximum input entropy among all channel outputs in $\Sigma_{q,R}^{m}$ is
 \begin{align*}
\begin{split}
\max_{\bfy \in \Sigma_{q,R}^{m}}{\mathsf{H}^{\mathsf{In}}_{1\textrm{-}\mathsf{Del}} (\bfy)} &= \log(nq) -\frac{r}{nq}\left(\ceil{\frac{m}{R}}+1\right)\log\left(\ceil{\frac{m}{R}}+1\right) \\
&-\frac{R-r}{nq}\left(\floor{\frac{m}{R}}+1\right)\log\left(\floor{\frac{m}{R}}+1\right),
\end{split}
\end{align*}
 and it is attained only by balanced channel outputs. 
\end{lemma}

The next theorem states that the input entropy of the $1$-deletion channel is maximized by channel outputs with exactly $m$ runs (each of length one). 
\begin{theorem}\label{thm:max1del}
The maximum input entropy among all channel outputs in $\Sigma_{q}^{m}$ is
$$\mathsf{max}^{\mathsf{In}}_{1\textrm{-}\mathsf{Del}}(n)=\log(nq) -\frac{2m}{nq}$$
and is only attained by channel outputs with $m$ runs.
\end{theorem}
\begin{IEEEproof} 
Let $\bfy \in \Sigma_{q}^{m}$ and let $\cR\cL(\bfy) = (r_1,r_2,\ldots,r_{\rho(\bfy)})$. From Corollary \ref{thm:entropy-1-deletion}, it can be shown that,  
\begin{align*}
\mathsf{H}^{\mathsf{In}}_{1\textrm{-}\mathsf{Del}} (\bfy) 
& = \log(nq)+ \frac{m+\rho(\bfy)}{nq}\sum_{i=1}^{\rho(\bfy)}\frac{r_{i}+1}{m+\rho(\bfy)}\log\left( \frac{m+\rho(\bfy)}{r_{i}+1}\right) \\  
& \ \ \ -\frac{m+\rho(\bfy)}{nq}\log(m+\rho(\bfy)).%
\end{align*}%
Since $\displaystyle\sum_{i=1}^{\rho(\bfy)}\frac{r_{i}+1}{m+\rho(\bfy)} = 1$, Jensen's inequality implies that
$$
\mathsf{H}^{\mathsf{In}}_{1\textrm{-}\mathsf{Del}} (\bfy) \leq \log(nq) + \frac{m+\rho(\bfy)}{nq}\log\left(\frac{\rho(\bfy)}{m+\rho(\bfy)}\right). 
$$
 
Let $f(x):[1,m]\to \mathbb{R}$ be defined as $f(x) \triangleq \log(nq) + \frac{m+x}{nq}\log\left(\frac{x}{m+x}\right)$. The function $f$ is increasing w.r.t $x$ and
hence, for $r \in [1,m]$, $f(r) \leq f(m)$, and equality is attained if and only if $r=m$. Hence, among all sequences in $\Sigma_q^{m}$, channel outputs with $\rho(\bfy) = m$ have the maximum entropy $\mathsf{max}^{\mathsf{In}}_{1\textrm{-}\mathsf{Del}}(n) = \log(nq) -\frac{2m}{nq}.$
\end{IEEEproof}

\subsubsection{Minimum Input Entropy}
Similarly to the previous subsection, here we study the channel outputs $\bfy \in \Sigma_q^m$ that minimize the input entropy, where $m = n-1$. 

The next lemma simply follows from Claim~\ref{claim:HW} and by substituting $a = 1$ in Lemma~\ref{lm:max-sum-R-runs} and thus we omit the proof.

\begin{lemma}\label{lem:min_entropy}
 Let $R\in [m]$, the minimum input entropy among all channel outputs in  $\Sigma_{q,R}^{m}$ is
\begin{align*}
\min_{\bfy \in \Sigma_{q,R}^{m}} {\mathsf{H}^{\mathsf{In}}_{1\textrm{-}\mathsf{Del}} (\bfy)} &=\log(nq) -\frac{(m-R+2)\log(m-R+2) + 2(R-1)}{nq}
\end{align*}
and it is attained only by skewed channel outputs. 
\end{lemma}
 
The next theorem states that the channel outputs that minimize input entropy of the $1$-deletion channel have a single run. This extends the results from~\cite{ABC18} to the non-binary case. 
\begin{theorem}\label{thm:min1del}
The minimum input entropy among all channel outputs in $\Sigma_{q}^{m}$ is
\begin{align*}
    \mathsf{min}^{\mathsf{In}}_{1\textrm{-}\mathsf{Del}}(n)=\log(nq) -\frac{\log(n)}{q}
\end{align*}
and is only attained by channel outputs having a single run.
\end{theorem}
\begin{IEEEproof}
According to Lemma~\ref{lem:min_entropy}, we have that
$$\min_{ \bfy \in \Sigma_{q,R}^{m}}{\mathsf{H}^{\mathsf{In}}_{1\textrm{-}\mathsf{Del}} (\bfy)}   =  \log(nq) -\frac{(m-R+2)\log(m-R+2) + 2(R-1)}{nq}.$$
It is easy to verify that the input entropy decreases as $R$ decreases. Therefore,
$\mathsf{min}^{\mathsf{In}}_{1\textrm{-}\mathsf{Del}}(n) = \log(nq) -\frac{\log(n)}{q} $ and it is attained if and only if $\rho(\bfy) = 1$.
\end{IEEEproof}

\subsection{Extremum Values for the $1$-Insertion Channel}
This subsection studies the channel outputs $\bfy \in \Sigma_q^m$ that maximize and minimize the input entropy of the $1\textrm{-}\mathsf{Ins}$ channel, where $m\triangleq n+1$. 

The next lemma simply follows from Claim~\ref{claim:HW} and by substituting $a = 0$ in Lemma~\ref{lm:min-sum-R-runs} and thus we omit the proof. 

\begin{lemma}
If $R\in [m]$ and $r\equiv m \bmod R$, then 
\begin{align*}
\max_{\bfy \in \Sigma_{q,R}^{m}} {\mathsf{H}^{\mathsf{In}}_{1\textrm{-}\mathsf{Ins}} (\bfy)}= \log(m) - \frac{1}{m}\left(r \ceil{\frac{m}{R}}\log\ceil{\frac{m}{R}} + (R-r)\floor{\frac{m}{R}}\log\floor{\frac{m}{R}}\right),
\end{align*}
and the maximum is obtained only by balanced channel outputs.
\end{lemma} 
The next theorem states that the input entropy of the $1$-insertion channel is maximized by channel outputs with exactly $m$ runs (each of length one). 
\begin{theorem}\label{thm:max1ins}
The maximum input entropy among all channel outputs in $\Sigma_{q}^{m}$ is
$$\mathsf{max}^{\mathsf{In}}_{1\textrm{-}\mathsf{Ins}}(n)=\log(m)$$
and is only attained by channel outputs with $m$ runs.
\end{theorem}
\begin{IEEEproof}
    As $R\leq m,$ it can be verified that $\frac{1}{m}\left(r \ceil{\frac{m}{R}}\log\ceil{\frac{m}{R}} + (R-r)\floor{\frac{m}{R}}\log\floor{\frac{m}{R}}\right) = 0$ if and only if $R = m$. Therefore, $\mathsf{H}^{\mathsf{In}}_{1\textrm{-}\mathsf{Ins}} (\bfy) \leq \log(m)$, with equality if and only if $R = m$. 
\end{IEEEproof}

The next lemma simply follows from Claim~\ref{claim:HW} and by substituting $a = 0$ in Lemma~\ref{lm:max-sum-R-runs} and thus we omit the proof.
\begin{lemma}\label{lem:min_entropy_ins}
If $R\in [m]$, then 
\begin{align*}
\min_{\bfy \in \Sigma_{q,R}^{m}}{\mathsf{H}^{\mathsf{In}}_{1\textrm{-}\mathsf{Ins}} (\bfy)}=\log(m) -\frac{(m-R+1)\log(m-R+1)}{m}
\end{align*}
and the minimum is obtained only by the skewed channel outputs. 
\end{lemma}
The next theorem states that the channel outputs that minimize  input entropy of the $1$-insertion channel have a single run.
\begin{theorem}\label{thm:min1ins}
The minimum input entropy among all channel outputs in $\Sigma_{q}^{m}$ is
\begin{align*}
    \mathsf{min}^{\mathsf{In}}_{1\textrm{-}\mathsf{Ins}}(n)= 0
\end{align*}
and is only attained by channel outputs having a single run.
\end{theorem}
\begin{IEEEproof}
According to Lemma~\ref{lem:min_entropy_ins}, we have that
$$\min_{\bfy \in \Sigma_{q,R}^{m}}{\mathsf{H}^{\mathsf{In}}_{1\textrm{-}\mathsf{Ins}} (\bfy)}=\log(m) -\frac{(m-R+1)\log(m-R+1)}{m}.$$
It is easy to verify that the input entropy decreases as $R$ decreases. Therefore,
$\mathsf{min}^{\mathsf{In}}_{1\textrm{-}\mathsf{Ins}}(n)= 0 $ and it is attained if and only if $\rho(\bfy) = 1$.
\end{IEEEproof}

\begin{figure*}[!ht]
\center
\includegraphics[width=.85\linewidth, height =20ex]{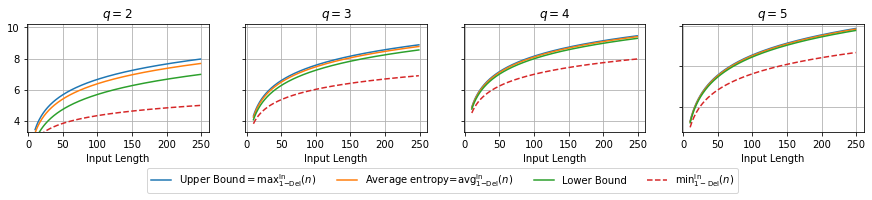}
 \caption{The minimum, maximum, average values and a lower bound on the average of the input entropy of the 1-deletion channel. } \label{fig:bounds}
\end{figure*}

\subsection{Average Values for the $1$-insertion and $1$-deletion channels}\label{sec:ave}
Let $N_{n,r,q}$ be the number of runs of length $r$ over all the sequences in $\Sigma_{q}^{n}$, where $n, q \in \mathbb{Z}_{\geq 1}$ and $r \in [n]$. We derive the value of $N_{n,r,q}$ in the following lemma.  
\begin{lemma}
    For $n, q \in \mathbb{Z}_{\geq 1}$ and  $r \in [n]$, it holds that 
\begin{align*}
        N_{n,r,q} = \begin{cases}
            q & r = n, \\
            2q(q-1) & r = n-1, \\
            (q-1)q^{n-r-1}\left((q-1)(n-r+1)+2 \right) & 1 \leq r \leq n-2. 
        \end{cases}
\end{align*}
\end{lemma}
\begin{IEEEproof}
    For $r = n$, it is easy to observe that $ N_{n,n,q} = q$. Now let $1 \leq  r \leq  n-1$. Let $N_1$ be the number of runs that start at the first position of any sequence. Similarly, let $N_2$ be the number of runs that end at the last position of any sequence. Then, $N_1 = N_2 = q(q-1)q^{n-(r+1)}$. Therefore, for $r = n-1$, we have that $N_{n,n-1,q} =  2q(q-1)$. Now let $1 \leq  r \leq  n-2$. Let $N_3$ be the number of runs that neither start at the first position nor end at the last position of any sequence. Then, $N_3 = \big((n-(r+2))+1\big)q(q-1)^2q^{n-(r+2)}.$ Note that $N_{n,r,q} = N_1 + N_2 + N_3$. After some simplification, it follows that
    \begin{align*}
        N_{n,r,q} = \begin{cases}
            q & r = n, \\
            2q(q-1) & r = n-1, \\
            (q-1)q^{n-r-1}\left((q-1)(n-r+1)+2 \right) & 1 \leq r \leq n-2. 
        \end{cases}
    \end{align*}
\end{IEEEproof}
We now derive the expressions for the average input entropies for the $1\textrm{-}\mathsf{Del}$ and $1\textrm{-}\mathsf{Ins}$ channels.  

\begin{lemma} \label{lm:avg}
For $n, q \in \mathbb{Z}_{\geq 1}$, it holds that 
\begin{align*}
&\mathsf{avg}^{\mathsf{In}}_{1\textrm{-}\mathsf{Del}}(n) = \log(nq) - \frac{\sum_{r=1}^{n-1}N_{n-1,r,q}(r+1)\log(r+1)}{nq^{n}},\\
&\mathsf{avg}^{\mathsf{In}}_{1\textrm{-}\mathsf{Ins}}(n) = \log(n+1) - \frac{\sum_{r=1}^{n+1}N_{n+1,r,q}(r)\log(r)}{(n+1)q^{n+1}}.
\end{align*}
\end{lemma}
 \begin{IEEEproof}
The following equalities hold
\begin{align*}
\mathsf{avg}^{\mathsf{In}}_{1\textrm{-}\mathsf{Del}}(n) 
&=\log(nq) - \frac{1}{nq}\mathbb{E}_{\Sigma_{q}^{n-1}}\left[\sum_{i=1}^{\rho(\bfy)}(r_{i}+1)\log(r_{i}+1)\right]\\
&=\log(nq) - \frac{1}{nq^n}\sum_{\bfy\in\Sigma_{q}^{n-1}}\sum_{i=1}^{\rho(\bfy)}(r_{i}+1)\log(r_{i}+1) \\
&= \log(nq) - \frac{1}{nq^{n}}\sum_{r=1}^{n-1}N_{n-1,r,q}(r+1)\log(r+1)\\
\end{align*}
The value of $\mathsf{avg}^{\mathsf{In}}_{1\textrm{-}\mathsf{Ins}}(n)$ can be evaluated similarly.
\end{IEEEproof}

Clearly, $\mathsf{min}^{\mathsf{In}}_{1\textrm{-}\mathsf{Del}}(n) \le \mathsf{avg}^{\mathsf{In}}_{1\textrm{-}\mathsf{Del}}(n)\le \mathsf{max}^{\mathsf{In}}_{1\textrm{-}\mathsf{Del}}(n)$ and $\mathsf{min}^{\mathsf{In}}_{1\textrm{-}\mathsf{Ins}}(n) \le \mathsf{avg}^{\mathsf{In}}_{1\textrm{-}\mathsf{Ins}}(1)\le \mathsf{max}^{\mathsf{In}}_{1\textrm{-}\mathsf{Ins}}(n)$ for $n,q \in \mathbb{Z}_{\geq 1}$. 

To improve the lower bounds on the average entropies which are better than the minimum values, note that if $r\geq1$, then $r\geq\log(r+1)$. Hence by Lemma~\ref{lm:avg}, it can be shown that
\begin{align*}
\mathsf{avg}^{\mathsf{In}}_{1\textrm{-}\mathsf{Del}}(n)   &\geq  \log(nq) - \frac{1}{n}\left(\frac{2n}{q-1}-\frac{n^2-n}{q^{n+1}}+\frac{2q^2-2q^{n+2}}{(q-1)^2 q^{n+1}}\right)
\end{align*}
and 
\begin{align*}
\mathsf{avg}^{\mathsf{In}}_{1\textrm{-}\mathsf{Ins}}(n)   &\geq    \log(nq)    +    \frac{1}{n+1}    \left( \frac{n^2}{q^{n+2}}  -  \frac{n(2q^{n+2}  -  q  +  1)}{(q  -  1)q^{n+2}}  +  \frac{2(q^n  -  1)}{(q  -  1)^2q^n}  \right) .
\end{align*}
The bounds for $\mathsf{avg}^{\mathsf{In}}_{1\textrm{-}\mathsf{Del}}(n)$ are presented in Figure~\ref{fig:bounds}.

\section{Extremum Values for $k \geq 2$}
\label{sec:extremum2}

In the following lemma, we analyze how the embedding number $\omega_\bfy(\bfx)$ is affected when a bit $\alpha$ is appended at the beginning of both $\bfx$ and $\bfy$.
 
\begin{lemma}\label{alpha}
Let $\bfy \in \Sigma_2^{m}$ and $\bfx \in \Sigma_2^{n}$ with $m \leq n$. For $\alpha \in \Sigma_2$, $\bfy' = \alpha\circ\bfy$ and $\bfx' = \alpha \circ \bfx$ it holds that   
$$\omega_{\bfy'}(\bfx') = \omega_{\bfy}(\bfx) + \sum_{i=1}^{k}\omega_{\bfy}(\bfx_{[i+1,n]})\cdot \mathbb{I}_{\alpha = x_i}.$$
\end{lemma}
\begin{IEEEproof}
It can be verified that
\begin{align}\label{DP-EW}
 {\omega_{\bfy}(\bfx)} &= \begin{cases} 
     \omega_{\bfy_{[2,m]}}(\bfx_{[2,n]}) + \omega_{\bfy}(\bfx_{[2,n]}) & y_1 = x_{1}, \\
     \omega_{\bfy}(\bfx_{[2,n]})& y_1 \neq x_{1}.\\ 
  \end{cases} 
\end{align}
Therefore, $\omega_{\bfy'}(\bfx') = \omega_{\bfy}(\bfx) + \omega_{\bfy'}(\bfx)$. Using \eqref{DP-EW} recursively, it follows that $\omega_{\bfy'}(\bfx) = \sum_{i=1}^{k}\omega_{\bfy}(\bfx_{[i+1,n]})\cdot\mathbb{I}_{\alpha = x_i}$.
\end{IEEEproof}

\begin{remark}
    As highlighted in Lemma~\ref{correction}, the characterization of embedding numbers differs slightly from that presented in \cite{ABC18}. Consequently, determining the exact value of $\mathsf{min}^{\mathsf{In}}_{2\textrm{-}\mathsf{Del}}(n)$, as well as identifying the sequences that attain this minimum, becomes more challenging under the corrected characterization due to the involvement of alternating segments. Accordingly, we re-establish the result in Theorem~\ref{thm:min-2del} to reflect the corrected characterization.
\end{remark}

\subsection{Minimum Input Entropy for $2$-Deletion Channel} 
In this section, we find the minimum value of the input entropy of $2\textrm{-}\mathsf{Del}$ Channel, and the sequences that attain this entropy among all sequences in $\Sigma_2^{m}$,  where $m \triangleq n-2$.

Before giving our main result of this section, we introduce some special sequences and analyse their properties. For $\bfy \in \Sigma_2^{m}$ with $\cR\cL(\bfy)=(r_1,r_2,\ldots, r_R)$ and $\cA\cL(\bfy)=(a_1,a_2,\ldots, a_A)$, let $x_\beta^{\bfy}, x_\gamma^{\bfy}, x_\delta^{\bfy} \in I_2(\bfy)$ be such that 

\begin{enumerate}
    \item $\cR\cL(\bfx_\beta^\bfy)\triangleq(1,1,r_1,r_2,\ldots, r_R)$,
    \item $\cR\cL(\bfx_\gamma^\bfy)\triangleq(2,r_1,r_2,\ldots, r_R)$,
    \item $\cR\cL(\bfx_\delta^\bfy)\triangleq(1,r_1+1,r_2,\ldots, r_R)$.
\end{enumerate}  
\begin{remark}
    Note that $x_\beta^{\bfy} \in y_1 \circ I_2(\bfy_{[2,m]})$ and $x_\gamma^{\bfy}, x_\delta^{\bfy} \in \overline{y}_1 \circ I_1(\bfy)$ . 
\end{remark}
The next two corollaries follow from Lemma~\ref{correction} and Lemma~\ref{alpha}.
\begin{corollary}\label{exc}
 For $\bfy \in \Sigma_2^{m}$ with $\cR\cL(\bfy)=(r_1,r_2,\ldots, r_R)$ and $\cA\cL(\bfy)=(a_1,a_2,\ldots, a_A)$, it holds that: 
 \begin{enumerate}
 \item $\omega_{\bfy}(\bfx_\beta^{\bfy}) = 1+ \sum_{j = 1}^{f_1} a_j -(a_{f_1}-1)\cdot\{\mathbb{I}_{f_1\neq i}\}$, which is maximal when $f_1 = A$.
 \item $\omega_{\bfy}(\bfx_\gamma^{\bfy}) = 1$. 
 \item $\omega_{\bfy}(\bfx_\delta^{\bfy}) = r_1+1$,  which is maximal when $R = 1$. 
 \end{enumerate}
\end{corollary}
\begin{corollary}\label{app1}
Let $\alpha \in \Sigma_{2}, \bfy \in \Sigma_2^{m}, \bfx \in \Sigma_2^{n}$, and $\bfy' = \alpha\circ\bfy$, $\bfx' = \alpha \circ \bfx$. If $ \alpha \ne y_{1}$ or $y_1 \ne x_{1}$, then 
\begin{align*}
{\omega_{\bfy'}(\bfx')} = \begin{cases} 
     {\omega_{\bfy}(\bfx)} + \mathbb{I}_{\bfx = \bfx_\beta^\bfy} & \alpha \neq y_{1}, y_1 = x_{1}, \\
     2 \cdot {\omega_{\bfy}(\bfx)} + \mathbb{I}_{\bfx = \bfx_\gamma^\bfy} & \alpha \neq y_{1}, y_1 \neq x_{1}, \\ 
     {\omega_{\bfy}(\bfx)} + \mathbb{I}_{\bfx = \bfx_\delta^\bfy} & \alpha = y_{1}, y_1 \neq x_{1}.
  \end{cases}
\end{align*}
\end{corollary}

For the ease of exposition, we give the next two lemmas, which will be useful for deriving the main result of this section, i.e., the value of $\mathsf{min}^{\mathsf{In}}_{2\textrm{-}\mathsf{Del}}(n)$ and the sequences that attain this minimum value.

\begin{lemma} \label{double-del-diff}
    Let $\bfy \in \Sigma_2^{\ell}, y_1 \neq y_2, \cR\cL(\bfy_{[2,\ell]})\triangleq (r_1,r_2,\ldots,r_R)$ and $ \cA\cL(\bfy_{[2,\ell]})\triangleq (a_1,a_2,\ldots,a_A)$. Then, we have that 
    \begin{align*}
\mathsf{W}_{y_1\circ I_2(\bfy_{[2,\ell]})}(\bfy) &= 3\log(3) - 2 + 4\ell + 2\cdot W_{I_1(\bfy_{[2,\ell]})}(\bfy_{[2,\ell]}) \\
&\quad+ \left(\theta + 1 \right) \log \left( \theta + 1 \right)  -  \theta \log \left( \theta \right) + W_{y_2 \circ I_2(\bfy_{[3,\ell]})} (\bfy_{[2,l]}),
\end{align*}
where $\theta = 1+ \sum_{j = 1}^{f_1} a_j -(a_{f_1}-1)\cdot\{\mathbb{I}_{f_1\neq 1}\}$.
\end{lemma}

\begin{IEEEproof}
    For $y_1 \neq y_2$, $\bfx \in  I_2(\bfy_{[2,\ell]})$, from Corollary \ref{app1}, we know that 
\begin{align}\label{app1:diff}
    {\omega_{\bfy}(y_1\circ\bfx)} = \begin{cases} 
     {\omega_{\bfy_{[2,\ell]}}(\bfx)} + \mathbb{I}_{\bfx = \bfx_\beta^{\bfy_{[2,\ell]}}}  &  y_{2} = x_{1}, \\
     2\cdot {\omega_{\bfy_{[2,\ell]}}(\bfx)} + \mathbb{I}_{\bfx = \bfx_\gamma^{\bfy_{[2,\ell]}}}&  y_{2} \neq x_{1}.
\end{cases}
\end{align}
Since $I_2(\bfy_{[2,\ell]}) = y_2 \circ I_2(\bfy_{[3,\ell]})~\bigcup~\overline{y_2}\circ I_1(\bfy_{[2,\ell]})$, 
\begin{align*}
    \mathsf{W}_{y_1\circ I_2(\bfy_{[2,\ell]})}(\bfy) = \mathsf{W}_{y_1\circ y_2 \circ I_2(\bfy_{[3,\ell]})}(\bfy) + \mathsf{W}_{y_1\circ \overline{y_2}\circ I_1(\bfy_{[2,\ell]})}(\bfy).
\end{align*}
We now analyse the first term, 
\begin{align*}
    &\mathsf{W}_{y_1\circ y_2 \circ I_2(\bfy_{[3,\ell]})}(\bfy) \\
    &= \sum_{\bfx~\in~ y_2 \circ I_2(\bfy_{[3,\ell]})} \omega_{\bfy}(y_1\circ \bfx) \log (\omega_{\bfy}(y_1\circ \bfx))\\
    &= \sum_{\bfx~\in~ y_2 \circ I_2(\bfy_{[3,\ell]})} \left(\omega_{\bfy_{[2,\ell]}}(\bfx) + \mathbb{I}_{\bfx = \bfx_\beta^{\bfy_{[2,\ell]}}}\right) \log \left(\omega_{\bfy_{[2,\ell]}}(\bfx) + \mathbb{I}_{\bfx = \bfx_\beta^{\bfy_{[2,\ell]}}}\right)\\
    &= \left(\omega_{\bfy_{[2,\ell]}}(\bfx_\beta^{\bfy_{[2,\ell]}}) + 1 \right) \log \left(\omega_{\bfy_{[2,\ell]}}(\bfx_\beta^{\bfy_{[2,\ell]}}) + 1 \right)  -   \left(\omega_{\bfy_{[2,\ell]}}(\bfx_\beta^{\bfy_{[2,\ell]}}) \right) \log \left(\omega_{\bfy_{[2,\ell]}}(\bfx_\beta^{\bfy_{[2,\ell]}})  \right) \\
    &\hspace{0.4cm} + \sum_{\bfx~\in~ y_2 \circ I_2(\bfy_{[3,\ell]})} \omega_{\bfy_{[2,\ell]}}(\bfx) \log \left(\omega_{\bfy_{[2,\ell]}}(\bfx) \right) \\
    &= \left(\theta + 1 \right) \log \left( \theta + 1 \right)  -  \theta \log \left( \theta \right) + W_{y_2 \circ I_2(\bfy_{[3,\ell]})} (\bfy_{[2,l]}),
\end{align*}
where, from Corollary \ref{exc}, $\theta \triangleq \omega_{\bfy_{[2,\ell]}}(\bfx_\beta^{\bfy_{[2,\ell]}}) = 1+ \sum_{j = 1}^{f_1} a_j -(a_{f_1}-1)\cdot\{\mathbb{I}_{f_1\neq 1}\}$.\\
Next, we analyse the second term,
\begin{align*}
&\mathsf{W}_{y_1\circ \overline{y_2}\circ I_1(\bfy_{[2,\ell]})}(\bfy) \\
&= \sum_{\bfx~\in~\overline{y_2}\circ I_1(\bfy_{[2,\ell]})} \omega_{\bfy}(y_1\circ \bfx) \log (\omega_{\bfy}(y_1\circ \bfx))\\
    &= \sum_{\bfx~\in~\overline{y_2}\circ I_1(\bfy_{[2,\ell]})} \left(2\cdot\omega_{\bfy_{[2,\ell]}}(\bfx) + \mathbb{I}_{\bfx = \bfx_\gamma^{\bfy_{[2,\ell]}}}\right) \log \left(2\cdot\omega_{\bfy_{[2,\ell]}}(\bfx) + \mathbb{I}_{\bfx = \bfx_\gamma^{\bfy_{[2,\ell]}}}\right)\\
    &= 3\log(3) - 2 + \sum_{\bfx~\in~\overline{y_2}\circ I_1(\bfy_{[2,\ell]})} 2\cdot\omega_{\bfy_{[2,\ell]}}(\bfx) \log \left(2\cdot\omega_{\bfy_{[2,\ell]}}(\bfx) \right)\\
    &= 3\log(3) - 2 + 2\left(\sum_{\bfx~\in~\overline{y_2}\circ I_1(\bfy_{[2,\ell]})} \omega_{\bfy_{[2,\ell]}}(\bfx) \left(1+\omega_{\bfy_{[2,\ell]}}(\bfx) \right)\right)\\
    &= 3\log(3) - 2 + 2\cdot W_{I_1(\bfy_{[2,\ell]})}(\bfy_{[2,\ell]}) + 2\sum_{\bfx\in I_1(\bfy_{[2,\ell]})}\omega_{\bfy_{[2,\ell]}}(\bfx)\\
    &= 3\log(3) - 2 + 4\ell + 2\cdot W_{I_1(\bfy_{[2,\ell]})}(\bfy_{[2,\ell]}).
\end{align*}
Therefore, 
\begin{align*}
\mathsf{W}_{y_1\circ I_2(\bfy_{[2,\ell]})}(\bfy) &= 3\log(3) - 2 + 4\ell + 2\cdot W_{I_1(\bfy_{[2,\ell]})}(\bfy_{[2,\ell]}) \\
&\quad+ \left(\theta + 1 \right) \log \left( \theta + 1 \right)  -  \theta \log \left( \theta \right) + W_{y_2 \circ I_2(\bfy_{[3,\ell]})} (\bfy_{[2,l]}),
\end{align*}
where $\theta = 1+ \sum_{j = 1}^{f_1} a_j -(a_{f_1}-1)\cdot\{\mathbb{I}_{f_1\neq 1}\}$.
\end{IEEEproof}

\begin{lemma} \label{double-del-same-A}
Let $\bfy \in \Sigma_2^{\ell}, y_1 = y_2, \cR\cL(\bfy_{[2,\ell]})\triangleq (r_1,r_2,\ldots,r_R)$ and $ \cA\cL(\bfy_{[2,\ell]})\triangleq (a_1,a_2,\ldots,a_A)$. Then, we have that 
    \begin{align*}
    \mathsf{W}_{y_1\circ \overline{y}_2\circ I_1(\bfy_{[2,\ell]})}(\bfy) &= (r_1+2)\log(r_1+2)  - (r_1+1)\log(r_1+1) \\
    &\quad+ W_{I_1(\bfy_{[2,\ell]})}(\bfy_{[2,\ell]}).
    \end{align*}
\end{lemma}
\begin{IEEEproof} Let $\bfx \in  I_2(\bfy_{[2,\ell]})$. Consider the following sum, 
    \begin{align*}
    &\mathsf{W}_{y_1\circ \overline{y}_2\circ I_1(\bfy_{[2,\ell]})}(\bfy) \\
    &= \sum_{\bfx~\in~\overline{y}_2\circ I_1(\bfy_{[2,\ell]})} \omega_{\bfy}(y_1 \circ \bfx)\log(\omega_{\bfy}(y_1 \circ \bfx))\\
    &= \sum_{\bfx~\in~\overline{y}_2\circ I_1(\bfy_{[2,\ell]})} \left(\omega_{\bfy_{[2,\ell]}}(\bfx) + \mathbb{I}_{\bfx = \bfx_\delta^{\bfy_{[2,\ell]}}}\right)\log\left(\omega_{\bfy_{[2,\ell]}}(\bfx) + \mathbb{I}_{\bfx = \bfx_\delta^{\bfy_{[2,\ell]}}}\right)\\
    &= \left(\omega_{\bfy_{[2,\ell]}}(\bfx_\delta^{\bfy_{[2,\ell]}}) + 1\right)\log\left(\omega_{\bfy_{[2,\ell]}}(\bfx_\delta^{\bfy_{[2,\ell]}}) + 1\right) - \left(\omega_{\bfy_{[2,\ell]}}(\bfx_\delta^{\bfy_{[2,\ell]}}) \right)\log\left(\omega_{\bfy_{[2,\ell]}}(\bfx_\delta^{\bfy_{[2,\ell]}}) \right) \\
    &\hspace{0.4cm} + \sum_{\bfx~\in~\overline{y}_2\circ I_1(\bfy_{[2,\ell]})} \omega_{\bfy_{[2,\ell]}}(\bfx)\log( \omega_{\bfy_{[2,\ell]}}(\bfx))\\
    &= (r_1+2)\log(r_1+2) - (r_1+1)\log(r_1+1) \\
    &\hspace{0.4cm}+ \sum_{\bfx_{[2,l+2]}~\in~ I_1(\bfy_{[2,\ell]})} \omega_{\bfy_{[2,\ell]}}(\bfx_{[2,l+2]})\log( \omega_{\bfy_{[2,\ell]}}(\bfx_{[2,l+2]}))\\
    &= (r_1+2)\log(r_1+2) - (r_1+1)\log(r_1+1) + W_{I_1(\bfy_{[2,\ell]})}(\bfy_{[2,\ell]}).
\end{align*}
\end{IEEEproof}

Let $0^\ell, 1^\ell \in \Sigma_{2}^\ell$ be the constant sequences of length $\ell$, i.e., $\rho(0^\ell) = \rho(1^\ell) = 1$. In the next lemma, we show that $\bfy = 0^m, 1^m$ minimize $\mathsf{H}^{\mathsf{In}}_{2\textrm{-}\mathsf{Del}} (\bfy)$ among all sequences in $\Sigma_2^m$, where $m \triangleq n-2$. 

\begin{theorem}\label{thm:min-2del}
The value $\mathsf{H}^{\mathsf{In}}_{2\textrm{-}\mathsf{Del}} (\bfy)$ is minimized only by $0^m, 1^m$, and 
$$
\mathsf{min}^{\mathsf{In}}_{2\textrm{-}\mathsf{Del}}(n) = 2 + \dfrac{3}{4}\log\binom{m}{2}-\dfrac{1}{2}\log(m+1).
$$
\end{theorem}
\begin{IEEEproof} We prove the lemma by induction on $m$. Let
\begin{align}\label{ind1}
    \bfg_m \triangleq \underset{\bfy\in \Sigma_2^{m}}{\argmax}~ \mathsf{W}_{y_1\circ I_2(\bfy_{[2:m]})}(\bfy).
\end{align}
For the base case, it can be verified that $\bfg_2 = \{00,11 \}$. For $\ell<m$, assume that $\bfg_{\ell-1} = 0^{\ell-1}$. We will now show that $\bfg_{\ell} = 0^{\ell}$. Let $\bfy \in \Sigma_2^{\ell}, \cR\cL(\bfy_{[2,\ell]})\triangleq (r_1,r_2,\ldots,r_R)$ and $ \cA\cL(\bfy_{[2,\ell]})\triangleq (a_1,a_2,\ldots,a_A)$. Let $\bfx \in  I_2(\bfy_{[2,\ell]})$. Consider the next two cases: \\

\noindent\textbf{Case I} - $y_1 \neq y_2$: From Lemma \ref{double-del-diff}, we know that 
\begin{align*}
\mathsf{W}_{y_1\circ I_2(\bfy_{[2,\ell]})}(\bfy) &= 3\log(3) - 2 + 4\ell + 2\cdot W_{I_1(\bfy_{[2,\ell]})}(\bfy_{[2,\ell]}) \\
&\quad+ \left(\theta + 1 \right) \log \left( \theta + 1 \right)  -  \theta \log \left( \theta \right) + W_{y_2 \circ I_2(\bfy_{[3,\ell]})} (\bfy_{[2,l]}),
\end{align*}
where $\theta = 1+ \sum_{j = 1}^{f_1} a_j -(a_{f_1}-1)\cdot\{\mathbb{I}_{f_1\neq 1}\}$.
It can be readily verified that $\left(a + 1 \right) \log \left( a + 1 \right)  -  a \log \left( a \right)$ is an increasing function with respect to $a$. From Corollary \ref{exc}, we know that $ \theta = \left(1+ \sum_{j = 1}^{f_i} a_j -(a_{f_i}-1)\cdot\{\mathbb{I}_{f_i\neq i}\}\right)$ is maximized when $f_1= A$. Note that for $\bfy_{[2:\ell]} = 0^{\ell-1}, f_1 = A$. Hence, from the induction assumption, i.e. $\bfg_{\ell-1} = 0^{\ell-1}$, and Theorem \ref{thm:min1del},  it follows that 
\begin{align*}
    \underset{y_1\neq y_2, \bfy_{[2:\ell]}\in \Sigma_2^{\ell-1}}{\argmax}~\mathsf{W}_{y_1\circ I_2(\bfy_{[2,\ell]})}(\bfy) = 1\circ0^{\ell-1}.
\end{align*}

\noindent\textbf{Case II} - $y_1 = y_2$: 
 Since $I_2(\bfy_{[2,\ell]}) = y_2 \circ I_2(\bfy_{[3,\ell]})~\bigcup~\overline{y_2}\circ I_1(\bfy_{[2,\ell]})$, 
\begin{align*}
    \mathsf{W}_{y_1\circ I_2(\bfy_{[2,\ell]})}(\bfy) = \mathsf{W}_{y_1\circ y_2 \circ I_2(\bfy_{[3,\ell]})}(\bfy) + \mathsf{W}_{y_1\circ \overline{y_2}\circ I_1(\bfy_{[2,\ell]})}(\bfy).
\end{align*}
We will now show that both of the sums are maximized by $0^{\ell}$. From Lemma \ref{double-del-same-A}, we know that 
\begin{align*}
        \mathsf{W}_{y_1\circ \overline{y}_2\circ I_1(\bfy_{[2,\ell]})}(\bfy) &= (r_1+2)\log(r_1+2)  - (r_1+1)\log(r_1+1) \\
    &+ W_{I_1(\bfy_{[2,\ell]})}(\bfy_{[2,\ell]}).
\end{align*}
We know that $\left(a + 1 \right) \log \left( a + 1 \right)  -  a \log \left( a \right)$ is an increasing function with respect to $a$. Therefore, $\left(r_1 + 2 \right) \log \left( r_1 + 2 \right)  -  \left(r_1+1\right) \log \left( r_1+1 \right)$ is maximized when $r_1 = \ell-1$. Note that for $\bfy_{[2:\ell]} = 0^{\ell-1}, r_1 = \ell -1$. Therefore, from Theorem \ref{thm:min1del},  it follows that 

\begin{align*}
    \underset{y_1 = y_2, \bfy_{[2:\ell]}\in \Sigma_2^{\ell-1}}{\argmax}~\mathsf{W}_{y_1\circ \overline{y}_2\circ I_1(\bfy_{[2,\ell]})}(\bfy) = 0^\ell.
\end{align*}
Now let
\begin{align*}
    \cW(\bfy_{[2,\ell]}) \triangleq \mathsf{W}_{y_1\circ y_2\circ I_2(\bfy_{[3:\ell]})}(\bfy) - \mathsf{W}_{ y_2\circ I_2(\bfy_{[3:\ell]})}(\bfy_{[2,\ell]}).
\end{align*}
In the Appendix, we show that $\underset{\bfy_{[2,\ell]} \in \Sigma_2^{\ell-1}}{\argmax}~\cW(\bfy_{[2,\ell]}) = 0^{\ell-1}$. Since $\bfg_{\ell-1} = 0^{\ell-1}$, it follows that
$$\underset{y_1 = y_2, \bfy_{[2:\ell]}\in \Sigma_2^{\ell-1}}{\argmax}~\mathsf{W}_{y_1\circ y_2\circ I_2(\bfy_{[3:\ell]})}(\bfy) = 0^{\ell}.$$
Therefore, 
$$\underset{y_1 = y_2, \bfy_{[2:\ell]}\in \Sigma_2^{\ell-1}}{\argmax}~\mathsf{W}_{y_1\circ I_2(\bfy_{[2:\ell]})}(\bfy) = 0^{\ell}.$$
The following result can be obtained by a manual comparison of these two cases,
$$\underset{\bfy \in \Sigma_2^{\ell}}{\argmax}~\mathsf{W}_{y_1\circ I_2(\bfy_{[2,\ell]})}(\bfy)=0^\ell.$$
From \eqref{DP-EW}, we have that ${\omega_{\bfy}(\overline{y}_1\circ\bfx)} = {\omega_{\bfy}(\bfx)}$. Therefore, from Theorem \ref{thm:min1del} and  Claim~\ref{claim:HW}, it can be deduced that
\begin{align}
    &\underset{\bfy \in \Sigma_2^{\ell}}{\argmax}~\mathsf{W}_{\overline{y_1}\circ I_1(\bfy)}(\bfy)=0^{\ell}.\label{ind2}
\end{align}
Thus, upon collecting the two sums together, we get  
$$\underset{\bfy \in \Sigma_2^{\ell}}{\argmax}~\mathsf{W}_{I_2(\bfy)}(\bfy)=\underset{\bfy \in \Sigma_2^{\ell}}{\argmax}\left\{\mathsf{W}_{y_1\circ I_2(\bfy_{[2,\ell]})}(\bfy) + \mathsf{W}_{\overline{y}_1\circ I_1(\bfy)}(\bfy)\right\}=0^\ell.$$
It can be verified that $\mathsf{W}_{I_2(\bfy)}(0^{\ell}) = \mathsf{W}_{I_2(\bfy)}(1^{\ell})$, and hence the result follows. 
\end{IEEEproof}

\subsection{Minimum Input Entropy for the $k$-Insertion Channel}\label{sec:minkIns}
In this section, we study the channel outputs $\bfy \in \Sigma_q^m$ that minimize the input entropy of $\kIns$ Channel, where $m \triangleq n+k$ and $k \geq 2$. 

\begin{theorem}\label{min-kIns}
The value $\mathsf{H}^{\mathsf{In}}_{\kIns} (\bfy)$ is minimized only by $0^m, 1^m$, and 
$$
\mathsf{min}^{\mathsf{In}}_{\kIns}(n) = 0.
$$
\end{theorem}
\begin{IEEEproof}
    Let $\bfy \in \Sigma_2^{m}$. It is known that $\sum_{\bfx \in D_k(\bfy)} \omega_{\bfx}(\bfy) = \binom{m}{k}$. 
    For $t \geq 1$, let $A_t \triangleq \{(a_1,a_2,\ldots, a_t) : a_i \geq 1~\forall~i \in [t], \sum_{i = 1}^{t} a_i = \binom{m}{k}, a_i \leq a_j~\text{if}~i<j\}$. Let $(a_1,a_2,\ldots,a_t), (a_1', a_2',\ldots a_t') \in A_t$ such that
\begin{align*}
a_i' = \begin{cases} 
      a_i+1 & i = \ell \\
      a_i-1 & i = s\\
      a_i & \textrm{otherwise.} 
 \end{cases}    
\end{align*}
Next, consider the difference, $\sum_{i=1}^{t}(a_i')\log(a_i') - \sum_{i=1}^{R}(a_i)\log(a_i)$
\begin{align*}
& = g(a_{\ell}+1) - g(a_s), 
\end{align*}
where $g(r) \triangleq (r+1)\log(r+1) - r\log(r)$, and note that $g$ is an increasing function w.r.t. $r$. Therefore, 
\begin{align*}
\sgn\left(\sum_{i=1}^{t}(a_i')\log(a_i') - \sum_{i=1}^{R}(a_i)\log(a_i)\right) = \sgn(a_{\ell}+1-a_{s}),
\end{align*}
where $\sgn$ denotes the signum function.  
Thus, it follows that 
\begin{align*}
\underset{(a_1,a_2,\ldots,a_t) \in A_t}{\argmax} \sum_{i = 1}^{t} a_i\log(a_i) = \left(1,1,\ldots 1, \binom{m}{k}-t+1\right). 
\end{align*}
Therefore, 
\begin{align*}
\mathsf{H}^{\mathsf{In}}_{\kIns} (\bfy) &= \log  \binom{m}{k} -\frac{\mathsf{W}_{D_k(\bfy)}(\bfy)}{\binom{m}{k}}\\
&\geq \log  \binom{m}{k} -\frac{\left(\binom{m}{k} - \vert D_{k}(\bfy)\vert +1\right) \log \left(\binom{m}{k} - \vert D_{k}(\bfy)\vert +1\right)}{\binom{m}{k}}, 
\end{align*}
where the right hand side is equal to $0$ if and only if $\vert D_k(\bfy)\vert = 1$. Note that $\vert D_k(\bfy)\vert = 1$ if and only if $\bfy = 0^m$ or $1^m$. Therefore, for $\bfy \in \Sigma_2^{m}/\{0^m,1^m\}$, 
\begin{align*}
\mathsf{H}^{\mathsf{In}}_{\kIns} (\bfy) > 0.
\end{align*}
Furthermore, it can be easily verified that $\mathsf{H}^{\mathsf{In}}_{\kIns} (0^m) = \mathsf{H}^{\mathsf{In}}_{\kIns} (1^m) = 0$. 
\end{IEEEproof}

\section{Conclusion}\label{sec:conclusion}
In this work, we analyzed the conditional entropies of $k$-deletion and $k$-insertion channels and particularly for $k=1$ and $k=2$, we gave a more elaborate characterization by considering the runs and maximal alternating segments of a sequence. We derived the extremum values and the sequences that attain these extremum values for the input entropy of both $1$-deletion and $1$-insertion channels over arbitrary alphabet size. We also studied the average input entropies for $1$-deletion and $1$-insertion channels. We then derived the minimum values and the sequences that attain these minimum values for the input entropy of $2$-deletion channel and $k$-insertion channel over binary alphabet.

\subsection{Future Work and Open Problems}
Although this paper presents results of significant importance in characterizing the entropy of $k$-deletion and $k$-insertion channels, several open problems remain for future exploration:
\begin{itemize}
    \item \textbf{Maximal Input Entropy for $k=2$.} While we provide a complete characterization of the sequences that minimize input entropy, the precise identification of the sequences that \emph{maximize} input entropy for the $2$-deletion and $2$-insertion channels is still unresolved. Determining these sequences and the associated maximal entropy values remains an important open direction.
    
    \item \textbf{Extending Analysis to $k > 2$.} The extremization and average-case analysis for $k$-deletion and $k$-insertion channels with $k > 2$ remain largely open. A key challenge lies in the combinatorial explosion of the number of embeddings and the absence of a tractable description of the conditional probability distribution for general $k$. The structural techniques developed in this paper—based on runs and alternating segments—do not extend readily to higher values of $k$, suggesting the need for new analytical tools.
    
    \item \textbf{Non-Uniform Input Distributions.} All results in this work are obtained under the assumption of uniform transmission. Analyzing extremal and average conditional entropy values under general or capacity-achieving input distributions is an important next step. This line of inquiry is particularly relevant for understanding the channel capacity of $k$-deletion/insertion channels, where the optimal input distribution is currently unknown. 
\end{itemize}

\newpage
\begin{appendices}
\section{}
\begin{claim}
For $\bfy \in \Sigma_2^m$, let
\begin{align*}
    \cW(\bfy) \triangleq \mathsf{W}_{y_1\circ y_1\circ I_2(\bfy_{[2:m]})}(y_1 \circ \bfy) - \mathsf{W}_{ y_1\circ I_2(\bfy_{[2:m]})}(\bfy).
\end{align*}
Then we have that
\begin{align*}
    \underset{\bfy \in \Sigma_2^{m}}{\argmax}~\cW(\bfy) = 0^{m}~\text{and}~1^{m}.
\end{align*}
\end{claim}
\begin{IEEEproof}
Let $\cR\cL(\bfy)\triangleq (r_1,r_2,\ldots,r_R)$. Using Theorem \ref{thm:entropy-2-deletion} and Lemma \ref{correction}, we get that
\begin{align*}
    \cW(\bfy) &= \binom{r_1+3}{2}\log\binom{r_1+3}{2} - \binom{r_1+2}{2}\log\binom{r_1+2}{2} \\
    &\quad+\left(1+\theta\right)\log\left(1+\theta\right)  - \left(\theta\right)\log\left(\theta \right)\\
    &\quad+ (m-R+1)(r_1+2)\log\left(r_1+2\right) - (m-R+1)(r_1+1)\log\left(r_1+1\right)\\
    &\quad+ \sum_{i=2}^{R} \Big\{ (r_1+2)(r_i+1)\log\left((r_1+2)(r_i+1)\right) \Big\}\\
    &\quad-\sum_{i=2}^{R} \Big\{ (r_1+1)(r_i+1)\log\left((r_1+1)(r_i+1)\right) \Big\},
\end{align*}
where $ \theta = 1+ \sum_{j = 1}^{f_i} a_j -(a_{f_i}-1)\cdot\{\mathbb{I}_{f_i\neq i}\}$.

It can be readily verified that $\left(a + 1 \right) \log \left( a + 1 \right)  -  a \log \left( a \right)$ is an increasing function with respect to $a$. From Corollary \ref{exc}, we know that $ \theta = \left(1+ \sum_{j = 1}^{f_i} a_j -(a_{f_i}-1)\cdot\{\mathbb{I}_{f_i\neq i}\}\right)$ is maximized when $f_1= A$. Therefore, 
\begin{align*}    
\underset{\bfy \in \Sigma_2^{m}}{\argmax}~\left\{\left(1+\theta\right)\log\left(1+\theta\right)  - \left(\theta\right)\log\left(\theta\right)\right\} = 0^m.
\end{align*}
We now analyse the remaining terms of $\cW(\bfy)$. For ease of exposition, let 
\begin{align*}
   \cW'(\bfy) &\triangleq  (m-R+1)(r_1+2)\log\left(r_1+2\right) - (m-R+1)(r_1+1)\log\left(r_1+1\right) \\
    &\quad+ \sum_{i=2}^{R} \Big\{ (r_1+2)(r_i+1)\log\left((r_1+2)(r_i+1)\right) - (r_1+1)(r_i+1)\log\left((r_1+1)(r_i+1)\right) \Big\}.
\end{align*}
We now assume to the contrary that for the sequence $\bfy'\in\Sigma_2^m$ with $\cR\cL(\bfy')\triangleq (r_1+1,r_2,\ldots, r_j-1, \ldots, r_R)$, the following holds: 
\begin{align*}
    &\cW'(\bfy)+ \binom{r_1+3}{2}\log\binom{r_1+3}{2} - \binom{r_1+2}{2}\log\binom{r_1+2}{2}\\ 
    >\quad &\cW'(\bfy') + \binom{r_1+4}{2}\log\binom{r_1+4}{2} - \binom{r_1+3}{2}\log\binom{r_1+3}{2}.
\end{align*}

First consider the difference, $\cW'(\bfy') - \cW'(\bfy)$,
\begin{align*}
\cW'(\bfy') - \cW'(\bfy) &=(m-R+1)(r_1+3)\log\left(r_1+3\right) - (m-R+1)(r_1+2)\log\left(r_1+2\right) \\
&\quad+ \sum_{i=2}^{R} \Big\{ (r_1+3)(r_i+1)\log\left((r_1+3)(r_i+1)\right) - (r_1+2)(r_i+1)\log\left((r_1+2)(r_i+1)\right) \Big\} \\ 
&\quad- \Big\{ (r_1+3)(r_j+1)\log\left((r_1+3)(r_j+1)\right) - (r_1+2)(r_j+1)\log\left((r_1+2)(r_j+1)\right) \Big\} \\ 
&\quad+ \Big\{ (r_1+3)(r_j)\log\left((r_1+3)(r_j)\right) - (r_1+2)(r_j)\log\left((r_1+2)(r_j)\right) \Big\} \\ 
&\quad-\Bigg\{(m-R+1)(r_1+2)\log\left(r_1+2\right) - (m-R+1)(r_1+1)\log\left(r_1+1\right) \\
&\quad+ \sum_{i=2}^{R} \Big\{ (r_1+2)(r_i+1)\log\left((r_1+2)(r_i+1)\right) - (r_1+1)(r_i+1)\log\left((r_1+1)(r_i+1)\right) \Big\} \Bigg\}.
\end{align*}
After some simplification, it follows that 
\begin{align*}
\cW'(\bfy') - \cW'(\bfy)&=(m-R+1)\bigg\{(r_1+3)\log\left(r_1+3\right) - 2(r_1+2)\log\left(r_1+2\right) + (r_1+1)\log\left(r_1+1\right)\bigg\} \\
&\quad+ \sum_{i=2}^{R} (r_i+1)\Big\{ (r_1+3)\log\left(r_1+3\right) - 2(r_1+2)\log\left(r_1+2\right) + (r_1+1)\log\left(r_1+1\right) \Big\} \\ 
&\quad+ (r_1+2)\log(r_1+2) - (r_1+3)\log(r_1+3) + (r_j)\log(r_j) - (r_j+1)\log(r_j+1).\\
\end{align*}
Upon expanding the summation, we get 
\begin{align*}
\cW'(\bfy') - \cW'(\bfy)&= (r_1+2)\log(r_1+2) - (r_1+3)\log(r_1+3) + (r_j)\log(r_j) - (r_j+1)\log(r_j+1) \\
&\quad+ (2m-r_1)\bigg\{(r_1+3)\log\left(r_1+3\right) - 2(r_1+2)\log\left(r_1+2\right) + (r_1+1)\log\left(r_1+1\right)\bigg\}. \\
\end{align*}
We know that $(r + 1)\log(r + 1) - r\log(r)$ is an increasing function w.r.t. $r$. Therefore, we get the folloing lower bounded:
\begin{align*}
\cW'(\bfy') - \cW'(\bfy)&\geq (2m-r_1)\bigg\{(r_1+3)\log\left(r_1+3\right) - 2(r_1+2)\log\left(r_1+2\right) + (r_1+1)\log\left(r_1+1\right)\bigg\} \\
&\quad+ (r_1+2)\log(r_1+2) - (r_1+3)\log(r_1+3) + (m)\log(m) - (m+1)\log(m+1).
\end{align*}
Hence, the following expression:
\begin{align*}
   \cW'(\bfy') + \binom{r_1+4}{2}\log\binom{r_1+4}{2} - \binom{r_1+3}{2}\log\binom{r_1+3}{2} \\
   - \cW'(\bfy)+ \binom{r_1+3}{2}\log\binom{r_1+3}{2} - \binom{r_1+2}{2}\log\binom{r_1+2}{2}
\end{align*}
is lower bounded by
\begin{align*}
(2m-r_1)\bigg\{(r_1+3)\log\left(r_1+3\right) - 2(r_1+2)\log\left(r_1+2\right) + (r_1+1)\log\left(r_1+1\right)\bigg\} \\
+ (r_1+2)\log(r_1+2) - (r_1+3)\log(r_1+3) + (m)\log(m) - (m+1)\log(m+1)\\
+ \binom{r_1+4}{2}\log\binom{r_1+4}{2} - 2\binom{r_1+3}{2}\log\binom{r_1+3}{2} +\binom{r_1+2}{2}\log\binom{r_1+2}{2}.
\end{align*}
It can be verified that the above lower bound is a decreasing function w.r.t. $r_1$ and that at $r_1 = m$ it is greater than $0$. Therefore this a contradiction to our assumption. Hence, 
\begin{align*}
\underset{\bfy \in \Sigma_2^{m}}{\argmax}~\left\{
\cW'(\bfy)+ \binom{r_1+3}{2}\log\binom{r_1+3}{2} - \binom{r_1+2}{2}\log\binom{r_1+2}{2}\right\} = 0^m,
\end{align*}    
and thus $\underset{\bfy \in \Sigma_2^{m}}{\argmax}~\cW(\bfy) = 0^m$. Furthermore, it can be verified that $\cW(0^m) = \cW(1^m)$. 
\end{IEEEproof}
\end{appendices}

\end{document}